\documentclass[a4paper,fleqn]{mnras}
\usepackage{graphicx}
\usepackage[T1]{fontenc}
\usepackage{ae,aecompl}

\title[Abundances and kinematics of barium stars]{Chemical abundances and kinematics of barium stars}

\author[de Castro et al.]{D.~B.~de~Castro$^{1}$\thanks{E-mail:denise@on.br}, 
C.~B.~Pereira$^{1}$\thanks{E-mail:claudio@on.br}, 
F.~Roig$^{1}$\thanks{E-mail:froig@on.br}, 
E.~Jilinski$^{1,2}$\thanks{E-mail:jilinski@on.br},
N.~A.~Drake$^{1,3}$\thanks{E-mail:drake@on.br},
\newauthor C.~Chavero$^{1,4,5}$\thanks{E-mail:carochavero@gmail.com}
and 
J.~V.~Sales~Silva$^{1}$\thanks{E-mail:joaovictor@on.br}
\\ \\
$^{1}$Observat\'orio Nacional/MCTI, Rua Gen. Jos\'e Cristino, 77, 20921-400, 
Rio de Janeiro, Brazil\\
$^{2}$ Pulkovo Observatory, Russian Academy of Sciences, Pulkovskoye chaussee 65, 196140, Saint-Petersburg, Russia \\
$^{3}$ Saint Petersburg State University, Universitetski pr. 28, 198504, 
Saint Petersburg, Russia\\
$^{4}$ Observatorio Astronomico de C\'ordoba (OAC), Laprida 854, X5000BGR, C\'ordoba, Argentina\\
$^{5}$ Consejo Nacional de Investigaciones Cient\'{\i}ficas y T\'ecnicas (CONICET), Argentina}

\date{Accepted xxx. Received xxx; in original form xxxx}
\pubyear{2016}

\begin{document}

\label{firstpage}
\pagerange{\pageref{firstpage}--\pageref{lastpage}}

\maketitle

\begin{abstract} 
\par In this paper we present an homogeneous analysis of photospheric abundances
based on high-resolution spectroscopy of a sample of 182
barium stars and candidates. We determined atmospheric parameters,
spectroscopic distances, stellar masses, ages, luminosities and scale
height, radial velocities, abundances of the Na, Al,
$alpha$-elements, iron-peak elements, and s-process elements Y, Zr, La, Ce, 
and Nd. We employed the local-thermodynamic-equilibrium model atmospheres of
Kurucz and the spectral analysis code {\sc moog}.  We found that the
metallicities, the temperatures and the surface gravities for barium
stars can not be represented by a single gaussian distribution. 
The abundances of $alpha$-elements and iron peak elements are similar
to those of field giants with the same metallicity. Sodium presents some degree of
enrichment in more evolved stars that could be attributed to the NeNa
cycle. As expected, the barium stars show overabundance of the elements
created by the s-process. By measuring the mean heavy-element abundance pattern as
given by the ratio [s/Fe], we found that the barium stars present several degrees
of enrichment. We also obtained the [hs/ls] ratio by measuring 
the photospheric abundances of the Ba-peak and the Zr-peak elements.
Our results indicated that the [s/Fe] and the [hs/ls] ratios are strongly
anti-correlated with the metallicity. Our kinematical analysis showed
that 90\% of the barium stars belong to the thin disk
population. Based on their luminosities, none of the barium stars are
luminous enough to be an AGB star, nor to become self-enriched in the
s-process elements. Finally, we determined that the barium stars also
follow an age-metallicity relation.
\end{abstract}


\section{Introduction}

\par Barium stars are a family of peculiar red giants whose envelopes
exhibit overabundance of carbon, as well as of elements heavier than
iron.  First recognized by Bidelman \& Keenan (1951), these objects
show strong lines of the s-process elements, particularly Ba\,{\sc ii}
at 4554\,\AA\, and Sr\,{\sc ii} at 4077\,\AA , and CH, CN, and C$_{2}$
molecular bands. The elements heavier than iron are synthesized in the
interior of AGB (asymptotic giant branch) stars through the slow
neutron capture process (s-process), a neutron-capture chain starting
on Fe seed nuclei and synthesizing nuclides heavier than Fe located
along the valley of nuclear stability (Burbidge et al. 1957).  As a
result of the so-called ``third dredge-up'', the s-process enriched
material is brought to the surface of the AGB stars (Iben \& Renzini
1983; Busso et al. 2001).  Until the discovery of the binary nature of
the barium stars through radial velocity monitoring (McClure et
al. 1980), the origin of the s-process enhancements observed in these
stars presented a fundamental challenge to the stellar nucleosynthesis
theory and to post-main-sequence evolution, because barium stars are
not luminous enough to have undergone third dredge-ups on the AGB.
Since barium stars have low luminosities $M_{V}$$\sim0.0$ (Jaschek et
al. 1985), and stars on the thermally pulsing phase have luminosities
$M_{V}$$\sim-3.0$ to $-$5.0 (Iben \& Renzini 1983), the chemical
peculiarities observed in the barium stars can only be attributed to
mass transfer in the binary system from an AGB star (now the white
dwarf).  In fact, as showed by McClure (1983), 85\% of the barium
stars are binary stars. Those that are considered a single star can be
actually related to binary systems where the orbits are either
pole-on, or they are very eccentric and significant variations of
radial velocity occur only in a small phase range, like HD 123949
whose period is 9\,200 days and has an eccentricity of 0.972 (Pourbaix
et al. 2004).

\par The barium star phenomenon was also detected in other peculiar
binary systems, where the observed overabundances of the s-process
elements are due to the mass-transfer hypothesis. This is the case of
the CH stars (Luck \& Bond 1991), yellow-symbiotic stars (Smith et al
1996; Pereira \& Roig 2009), binary planetary nebulae such as Abel 35
and LoTr5 (Th\'evenin \& Jasniewicz 1997), and CEMP-s stars (Sivarani
et al. 2004).

\par Since the barium stars are warmer than the AGB stars with
temperatures below 3\,500\,K, they are free from the strong molecular
opacity caused by ZrO, CN, and C$_{2}$ bands absorption features,
which complicate the analysis and the abundance determinations based
on atomic lines. Therefore, the barium stars become very useful
targets where the abundances of several elements of the s-process can
be determined by measuring equivalent widths of their lines. This
probably explains why barium stars have attracted the attention of
several astrophysicists.  The number of barium stars investigated so
far is still small, and some stars have been investigated more than
once. Recent and earlier papers devoted to obtain the abundance
pattern in barium stars have been mainly interested to provide a
precise diagnostic of the s-process nucleosynthesis in these stars,
and also to set constraints for AGB models through the determination
of the [hs/ls]
\footnote{[hs/ls]$=\log{\rm(hs/ls)}_{\star}-\log{\rm (hs/ls)}_{\odot}$ where
[hs] and [ls] are the mean abundances of the s-elements at the Ba and Zr
 peaks, respectively.} ratio  (Tomkin \&
Lambert 1979; Sneden et al. 1981; Smith 1984; Antipova et al. 2003;
Liang et al. 2003, Allen \& Barbuy 2006; Smiljanic el al. 2007;
Pereira et al. 2011). Barium stars have been also used by Busso et
al. (2001) as a source sample of extrinsic Galactic AGB stars in a
study that aimed to compare the observed [hs/ls] ratios for several classes of
chemically peculiar stars (intrinsic and extrinsic) with that predicted by
nucleosynthesis models for AGB stars with different masses and
metallicities. In Busso et al. (2001), only 13 barium stars have been used to constrain
the [hs/ls] ratio for the extrinsic objects.  Later, Husti et
al. (2009) used a more extended sample from Allen \& Barbuy (2006) and
Smiljanic et al. (2007) to compare the observed abundances with their
AGB nucleosynthesis models for different masses, metallicities and
$^{13}$C efficiencies.

\par Among the chemically peculiar stars, the barium stars are
probably the largest sample. Han et al. (1995) estimated that the
number of the barium stars in our Galaxy may range from 800 to 22\,000
for stars brighter than $V$\,=\,10.0, and this includes both strong
and mild barium stars. The catalog of L\"u (1991) lists 389 barium
stars including ``certain'' and ``candidates'', and the earlier
publication of MacConnell et al. (1972) lists 241 barium stars
including ``certain'' and ``marginal''. The quantitative confirmation
of the overabundances of the s-process elements in such a large sample
of barium stars would help not only to better constrain the number of
``actual'' barium stars, but would also set important constraints for
the models of AGB nucleosynthesis. The former data would be useful to
compare with their theoretical birthrate, while the latter can be
achieved, for example, by determining the [hs/ls] ratio (among other
things), and would help to answer questions like how is the behavior
of the [hs/ls] ratio with metallicity and mass.  Recently, two results
appeared in the literature showing that the barium star studies may
still post interesting questions. The first one (Katime Santrich et
al. 2013) reports the discovery of two barium stars in the open
cluster NGC 5822, while the second one (Lebzelter et al. 2013) reports
the discovery of two barium stars in the Galactic bulge. Therefore,
investigating the metallicity of a large sample of barium stars, taken
from different sources, makes possible to probe the s-process
nucleosynthesis in different Galactic populations.

\par Aiming to assess the above issues, in 2007 we started a
high-resolution spectroscopic survey of a sample of barium stars using
the FEROS spectrograph (although a few objects had been already
observed between 1999 and 2001). The majority of the observed barium
stars were selected from MacConnell et al. (1972), but we refer the
reader to Section 2 for more details on Target Selection.  Previous
results from this survey include: the discovery of a new CH subgiant
star BD-03$^\circ$3668 (Pereira \& Drake 2011); the analysis of a
metal-poor barium star HD 10613 and the CH star BD+04$^\circ$2466
(Pereira \& Drake 2009); and the analysis of small sample of
metal-rich barium stars (Pereira et al. 2011).

\par Besides probing the enrichment of the s-process elements, the
high-resolution spectroscopic data of a large sample of barium stars
allows us to determine, among other things, the atmospheric
parameters. This makes possible to compare the distributions of
temperature, surface gravity, metallicity, and microturbulent velocity
to other studies already done for field giant stars. Then, we can
investigate if the barium giant stars have similar parameters to the
non-s-process enriched giant stars. The abundances of other elements
besides those of the s-process, such as sodium, aluminum,
$\alpha$-elements (Mg, Si, Ca, and Ti) and iron peak elements (Cr and
Ni) can also be obtained. This, in turn, would rise the question on
whether the abundances of barium stars are similar to those of the
field giants with equal metallicities. From the surface gravity and
temperature, and using evolutionary tracks and isochrones, the masses
and ages of the barium stars can be determined too. In addition, the
spectroscopic distance, obtained from the ionization equilibrium using
the abundances of the Fe\,{\sc i} and Fe\,{\sc ii} lines, allows us to
determine the luminosity of these stars and their scale
height. Luminosity is a key parameter to discuss the origin of
overabundances of the elements of the s-process. We can also obtain
the radial velocities of the stars in our sample, measuring Doppler
shifts of some absorption lines. Combining the radial velocities,
distances, and knowing the proper motions, it is possible to determine
the spatial velocities and to set a kinematical constraint for a given
population. The determination of the kinematics, the metallicity, and
the mean abundance of $\alpha$-elements provide another important
check on whether a correlation between abundances and kinematics holds
for the barium stars. Finally, we can also search for technetium lines
in the spectra of the more evolved stars.

\par In this work we observed and analyzed 182 stars among ``certain''
and ``candidate'' barium stars.  Our results are all based
on the measurement of equivalent widths.  The number of Fe\,{\sc i} and
Fe\,{\sc ii} absorption lines used was about 22\,000. To
determine the abundances of the other elements, the number of absorption
lines used was about 33\,000. In Sect. 2, we explain the selection 
criteria of our targets. Section 3 describes the observations. Section
4 presents, analyzes and discusses the results. Finaly, Sect. 5 
is devoted to the conclusions.

\section{Target selection}

\par The majority of the barium stars analyzed in this work were
selected from MacConnell et al. (1972), 109 stars out of 151 (72\%)
from his Table\,I (``Certain Ba\,{\sc ii} stars''), and 54 stars out
of 90 (60\%) from his Table\, II (``Marginal Ba\,{\sc ii} stars'').
Marginal barium stars from Table II of MacConnell et al. (1972) were
included in our study because we need to determine their heavy-element
abundance in order to reveal how strong is the degree of s-process
enrichment in these stars. This group of marginal stars represents an
interesting target to study the nucleosynthesis of the s-process,
since some metal-rich and s-process enriched stars have already been
found among them (Pereira et al. 2011).  Other barium stars were
selected from other sources.  We included in our analysis 12 out of 15
stars classified as ``Ba\,{\sc ii}'' by Bidelman (1981), and two stars
from Luck \& Bond (1991): BD+09$^\circ$2384 and HD 89638.

\par Some halo stars that are not included in the references above
were selected from two other sources that analyzed a large sample of
barium stars using previous Hipparcos astrometric and radial velocity
data. In particular, we selected four halo candidate stars,
BD-01$^\circ$302, HD 749, HD 88927, and HD 211211, from G\'omez et
al. (1997), and another candidate halo star, HD 187762, was taken from
Mennessier et al. (1997).  These latter authors list 20 candidate halo
stars, but we selected only one because several stars in their sample
have been recognized later as dwarf barium stars, or we already
selected them from MacConnell et al. (1972).  Although G\'omez et
al. (1997) and Mennessier et al. (1997) analyzed 318 and 296 stars,
respectively, taken from L\"u (1991), some of theses stars may not be
barium stars according to Jorissen et al. (1996).  This is because no
heavy-element overabundances have been found in some of them after the
high-resolution analysis of McWilliam (1990). For this reason, we
decided to select targets mainly from sources other than of L\"u
(1991).

\par Our sample is then constituted of 182 stars, among certain barium
stars and candidates.  They are listed in Table 1, where the last
column indicates the source from which they have been selected.
Throughout this work, we present our results following the same order
given in Table 1, except for Table 16 where we separate the stars
according to their kinematical properties.

\section{Observations} 

\par The high-resolution spectra of the barium stars analyzed in this
work were obtained with the FEROS (Fiberfed Extended Range Optical
Spectrograph) spectrograph (Kaufer et al. 1999), installed at the
1.52\,m and 2.2\,m telescopes of ESO at La Silla (Chile). Data was
obtained during several observing runs performed between 1999 and
2010. FEROS has a spectral resolving power of $R=48\,000$ with a
wavelength coverage from 3\,800\,{\AA} to 9\,200\,{\AA}. The spectra
were reduced with the MIDAS pipeline reduction package, consisting of
the following standard steps: CCD bias correction, flat-fielding,
spectrum extraction, wavelength calibration and correction of
barycentric velocity. Table 1 provides the log of observations and
information about the $V$-magnitude, spectral types, and exposure
times. The $V$-magnitudes and spectral types were taken from the
Simbad database. Figures 1 and 2 show the spectra of four very
s-process enriched barium stars.

\begin{figure} 
\includegraphics[width=10.5cm]{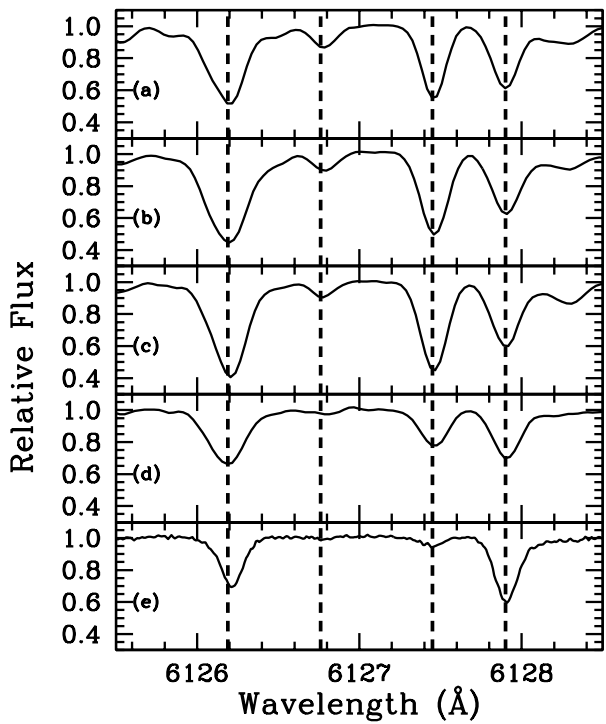}
\caption{Sample spectra of HD 24035 (a), HD 92626 (b), HD 123949, (c) and
HD 201824 (d) in comparison with a non s-process enriched giant HD 2114 (e).
Absorption lines due to the transitions of Ti\,{\sc i} 6126.19,
CN 6126.76, Zr\,{\sc i} 6127.48, and Fe\,{\sc i} 6127.91 are shown.
Dashed lines represent their rest wavelengths.}
\end{figure}

\begin{figure} 
\includegraphics[width=10.5cm]{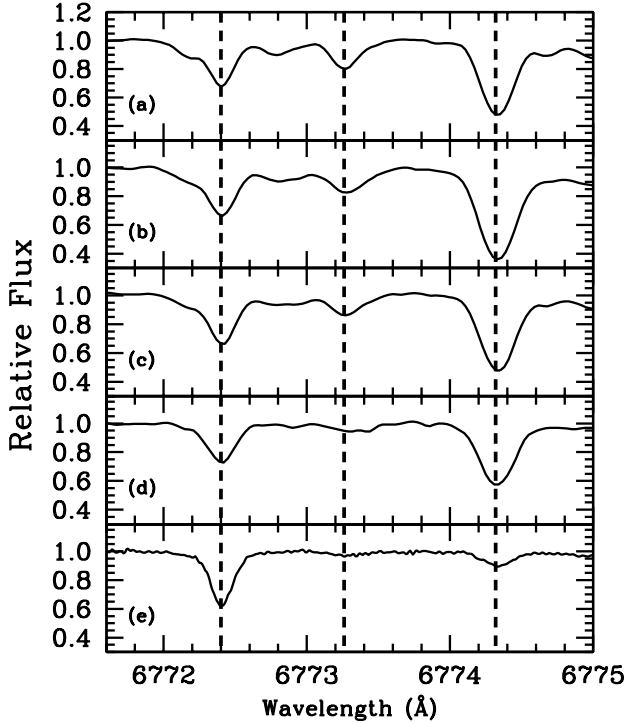}
\caption{Sample spectra of HD 24035 (a), HD 92626 (b), HD 123949, (c) and
HD 201824 (d) in comparison with a non s-process enriched giant HD 2114 (e).
Absorption lines due to the transitions of Ni\,{\sc i} 6772.32, CN 6773.26, 
and La\,{\sc ii} 6774.32 are shown. Dashed lines represent their rest wavelengths.}
\end{figure}

\section{Analysis and Results}

\subsection{Line selection and atmospheric parameters}

\par The absorption lines of Fe\,{\sc i} and Fe\,{\sc ii}, as well as
of other elements selected in this study, were the same used in
previous studies dedicated to the analysis of photospheric abundances
of some barium giants. The equivalent widths were measured by fitting
Gaussian profiles to the observed profiles.  The $\log gf$ values for
the Fe\,{\sc i} and Fe\,{\sc ii} lines were taken from Lambert et
al. (1996) and Castro et al. (1997).  In Table 2, we show our Fe\,{\sc
  i} and Fe\,{\sc ii} lines list used for the determination of the
atmospheric parameters. Only the lines with equivalent widths between
10\,m\AA\, and 150\,m\AA\, were used in the determination of
atmospheric parameters.

\par We recall that the determination of the stellar atmospheric
parameters, such as effective temperature ($T_{\rm eff}$), surface
gravity ($\log g$), microturbulence ($\xi$), and metallicity ([Fe/H])
(we use the notation [X/H]=$\log(N_{\rm X}/N_{\rm H})_{\star} -
\log(N_{\rm X}/N_{\rm H})_{\odot}$) are mandatory for the
determination of photospheric abundances.

\par The atmospheric parameters were determined using the LTE (local
thermodynamic equilibrium) model atmospheres of Kurucz (1993), and the
code of spectral analysis MOOG (Sneden 1973). The solution of
excitation equilibrium used to derive the effective temperature
($T_{\rm eff}$) was defined by a zero slope of the trend between the
iron abundances derived from Fe\,{\sc i} lines and the excitation
potential of the measured lines.  The microturbulent velocity ($\xi$)
was found by constraining the abundance, determined from individual
Fe\,{\sc i} lines, to show no dependence on reduced equivalent width
($W_{\lambda}/{\lambda}$).  As a by-product, this yields the
metallicity, [Fe/H], of the star.  The value of $\log g$ was
determined by means of the ionization balance using the assumption of
LTE, that is, the gravity is determined by forcing the Fe\,{\sc i} and
Fe\,{\sc ii} lines to yield the same iron abundance at the selected
effective temperature.  The adopted atmospheric parameters are given
in Table~3.  We also show the number of the Fe\,{\sc i} and Fe\,{\sc
  ii} lines employed for the determination.

\par The error in our adopted effective temperatures ($T_{\rm eff}$)
and microturbulent velocity ($\xi$) was set from the uncertainty in
the slope of the Fe\,{\sc i} abundance {\sl versus} excitation
potential and {\sl versus} $W_{\lambda}/{\lambda}$.  For the gravity,
the error in this parameter is estimated until the difference between
the mean abundances of Fe\,{\sc i} and Fe\,{\sc ii} differ by
1$\sigma$ of the standard deviation of the [Fe\,{\sc i}/H] mean value.
We noticed that the cooler stars of our sample (stars with
temperatures between 4\,100\,K and 4\,500\,K) give larger errors in
the temperature than the hotter stars (stars with temperatures between
5\,000\,K and 5\,400\,K). For these cooler stars we found a mean error
of 120$\pm$50\,K while for those hotter stars we found a mean error of
90$\pm$20\,K.  The reason for this difference is that cooler stars
gives higher abundance uncertainties since they have higher
uncertainty in the continuum placement which affects the measurement
of the equivalent widths.

\par The classification of a star as a barium star can be confirmed
only after the determination of the heavy-element abundance. Of the
182 stars initially selected, 13 were later rejected because their
mean s-process elements abundances in the notation [s/Fe] (see Section
4.3 for the definition of [s/Fe]) were similar to those of the field
giants.  These were BD-01$^\circ$302, HD 5322, HD 21980, HD 33409, HD
42700, HD 51315, HD 95345, HD 142491, HD 147136, HD 168986, HD 174204,
HD 211221, and HD 212484 (see discussion in in Section 5.2.1). These
rejected stars were replaced 11 metal-rich barium stars taken from
Pereira et al. (2011), and the stars HD 10613 (Pereira \& Drake 2009)
and HD 206983 (Junqueira \& Pereira 2001), previously analyzed by us.
The total sample of 182 barium stars was subject to a statistical
analysis of the temperatures, gravities, metallicities, microturbulent
velocities, masses, luminosities, abundances, and kinematics. These
results were compared to the same parameters for the field giants
taken from Mishenina et al. (2006), Luck \& Heiter (2007), Hekker \&
Melendez (2007), and Takeda et al. (2008).  The sample of field giants
excludes stars with $\log g$ $>$ 3.4, thus totalizing 1\,135 stars
used for comparison.

\par Figure 3(a-d) show the normalized histograms of temperature,
gravity, metallicity, and microturbulent velocity for the 182 barium
stars (red) and the 1\,135 field giants (black), together with the
corresponding gaussian fits to these distributions. In Table 4, we
list the values of the mean and standard deviation resulting from
these gaussian fits, as well as the simple mean of each histogram.

\par Figure 3(a) shows that the metallicity distribution of the barium
stars involves two gaussian components: a major peak corresponding to
the thin disk, and a smaller peak that corresponds to the thick
disk. The corresponding mean metallicities are $-$0.12$\pm$0.14 and
$-$0.49$\pm$0.09, respectively (Table 4). These values are similar to
those found by Schuster et al. (2006) in their photometric analysis of
disk and halo stars. These authors found a mean metallicity of
$-$0.16$\pm$0.14 for the thin disk, and a mean metallicity of
$-$0.55$\pm$0.18 for the thick disk.

\par Figure 3(b) shows the distribution of surface gravity for the
field giants and barium giants. The barium giants display three peaks
in the distribution: one peak at $\log g$\,=\,2.45, another peak at
$\log g$\,=\,1.60, and a third one at $\log g$\,=\,1.17. On the other
hand, the field giants are represented by only one broad gaussian
distribution, with a standard deviation of 0.42. Table 4 shows the
mean and tstandard deviations of the three distributions that fit
$\log g$ for the barium giants, and the mean and standard deviation of
the fit for the field giants.  The peak at $\log g$\,=\, 1.60 is due
to the presence of several barium giants with surface gravities
between 1.4 and 1.8. There are 34 (Table 3 and the star HD 206983;
Junqueira \& Pereira 2001) out of 182 barium stars in this range of
$\log g$, which represents 19\% of the sample.  This fraction is of
only 2.5\% (28 out of 1\,135 stars) in the sample of field giants. The
peak at $\sim$1.2 is due to the presence of six barium stars with
$\log g$ between 1.1 and 1.3, which represents 3.0\% of the
sample. The field giants in the same range of surface gravity
represents only 0.4\% (5 out of 1\,135 stars).

\par The distribution of surface gravity also reveals that the peak
for the barium giants is shifted by approximately 0.3 dex towards
lower values of $\log g$ in comparison to the peak of the field
giants. Differences in the analysis made by different authors, which
include the atomic data used for the metallicity and surface gravity
determination, the choice of model atmosphere grids, or even the
automated placement of the continuum, may account for differences of
up to 0.3--0.4 dex in the determination of $\log g$ and metallicity in
giant stars. However, evolutionary effects should also be
considered. Many stars from the samples of Mishenina et al. (2006),
Luck \& Heiter (2007), and Takeda et al. (2008) are clump giants, that
is, there are many stars with a $\log g$ between 2.0 and 3.0 (Figures
2 and 3a of Takeda et al. 2008).

\par Like the surface gravity, the peak of the distribution of
metallicity of barium giants also shows a shift of ($\sim$0.1 dex)
towards lower values in comparison to the field giants. The most
likely explanation for this difference is related to the systematic
differences in the analysis (see the end of this Section).  Most of
the previous studies of field giants used in this work to compare with
the barium stars, show that they are members of the thin disk
population.  Based on the spatial velocities, Luck \& Heiter (2007)
showed that their sample is dominated by stars from the thin
disk. Takeda et al. (2008) also shows that 97\% of the stars in their
sample also belong to thin disk.  In fact, as we show in Section 5.3
(Table 17), these two samples have values of the mean spatial
velocities $\langle$$U_{\rm LSR}$$\rangle$, $\langle$$V_{\rm
  LSR}$$\rangle$, $\langle$$W_{\rm LSR}$$\rangle$ and the
corresponding dispersions ($\sigma$$_{U}$, $\sigma$$_{V}$,
$\sigma$$_{W}$) very similar to those of the thin disk stars.

\par Figure 3(c) shows the distribution of temperature with two peaks
in the histogram for the barium giants. One peak at 4\,460\,K
corresponds to the more evolved stars that produce the peak at $\sim$
1.60 observed in the distribution of $\log g$ (Fig. 3(b)). The main
peak in the distribution of temperatures for the barium stars can be
fitted by a gaussian with similar mean and standard deviation as those
of the field giants.  Finally, concerning the microturbulent velocity,
Fig. 3(d) reveals no significant differences between the barium and
field giants.

\par In Figure 4, we show the correlations of the surface gravity and
the metallicity with respect to the temperature. In (a), $T_{\rm eff}$
decreases with lower $\log g$, while in (b), the metallicity shows no
trend with temperature.

\par In Figure 5, we show a diagram of $\log g$ {\sl versus}
metallicity for our samples of barium stars (red) and field giants
(black).  We also include in this diagram the field dwarf stars (blue)
analyzed by Edvardsson et al. (1993) and Reddy et al. (2003, 2006),
and the dwarf barium stars (green) analyzed by Luck \& Bond (1991),
Smith et al. (1993), North et al. (1994), Pereira \& Junqueira (2003),
Pereira (2005), Pereira \& Drake (2011), Allen \& Barbuy (2006), and
Liu et al. (2012). We can appreciate that number of known barium
dwarfs is still very small compared to the barium giants.  Unlike
these latter, that have been found over a wide range of metallicities,
the barium dwarfs have only been found down to
[Fe/H]$\le$$-$0.10. This point was highlighted by Pereira et
al. (2013), where we showed that metal-rich barium giants do exist and
probably belong to the thin-disk population, while barium dwarfs of
near solar metallicity have not been found yet.

\par Our sample of barium stars includes 33 stars, corresponding to
approximately 20\% of the sample, that have been already analyzed by
other authors and have atmospheric parameters determined from several
sources in the literature. Table 5 shows a comparison between these
previous results and our determinations in this study. As we can see,
our values for the atmospheric parameters do not differ significantly
from those in the literature, and the differences are related to the
different codes used for abundance determination, $gf$-values, choice
of lines, and, in some cases, the use of the differential analysis.

\begin{table*} 
\caption{Log of the observations and basic information of the stars.}
\begin{tabular}{cccccc}\hline
Star       & Date     & Exposure time & $V$ (mag)  & SpT$^{a}$ & Source \\
           &          &  (sec)       &          &          &        \\\hline
BD-08$^\circ$3194  & 08/04/2008 &  600 & 9.20  &  K1    &  I$^{b}$ \\
BD-09$^\circ$4337  & 08/04/2008 & 1200 & 9.70  &  K4    &\\
BD-14$^\circ$2678  & 19/02/2008 & 1200 & 9.82  &  K5    &\\
CD-27$^\circ$2233  & 15/10/2007 &  900 & 9.00  &  K2    &\\
CD-29$^\circ$8822  & 19/02/2008 & 1200 & 9.85  &  K0    &\\
CD-30$^\circ$8774  & 05/04/2007 & 1800 & 9.79  &  G5    &\\
CD-38$^\circ$585   & 25/22/2001 & 3600 & 9.94  &  K0    &\\
CD-42$^\circ$2048  & 17/10/2008 &  900 & 9.33  &  K5    &\\
CD-53$^\circ$8144  & 11/04/2008 &  900 & 9.18  &  K0    &\\
CD-61$^\circ$1941  & 19/02/2008 & 1200 & 9.29  &  M     &\\\hline
\end{tabular}

\par Table 1 is published in its entirety in the electronic edition 
of the Monthly Notices of the Royal Astronomical Society.
A portion is shown here for guidance regarding its form and content.
\end{table*}

\begin{table*}
\caption{Fe\,{\sc i} and Fe\,{\sc ii} lines used in our analysis. We also give
the reference for the log $gf$-values.}
\begin{tabular}{cccccccccc}\hline
Ion  & $\lambda$(\AA) & $\chi$(eV) & log $gf$ & Ref. & Ion & $\lambda$(\AA) & $\chi$(eV) & log $gf$ 
& Ref. \\\hline
Fe\,{\sc i} &  5242.49 & 3.63 & $-$0.97 & L &  Fe\,{\sc i} & 6170.51 & 4.80 & $-$0.38 & L\\
 & 5253.03 & 2.28 & $-$3.79 & L &  & 6173.34 & 2.22 & $-$2.88 & L\\
 & 5288.52 & 3.69 & $-$1.51 & L &  & 6187.99 & 3.94 & $-$1.57 & L\\
 & 5302.31 & 3.28 & $-$0.74 & L &  & 6200.32 & 2.60 & $-$2.44 & L\\
 & 5307.36 & 1.61 & $-$2.97 & L &  & 6213.43 & 2.22 & $-$2.48 & L\\
 & 5315.05 & 4.37 & $-$1.40 & L &  & 6230.72 & 2.56 & $-$1.28 & L\\
 & 5321.11 & 4.43 & $-$1.19 & L &  & 6254.26 & 2.28 & $-$2.44 & L\\
 & 5322.04 & 2.28 & $-$2.84 & L &  & 6265.13 & 2.18 & $-$2.55 & L\\
 & 5364.87 & 4.45 & $+$0.23 & L &  & 6311.50 & 2.83 & $-$3.23 & L\\
 & 5367.47 & 4.42 & $+$0.44 & L &  & 6322.69 & 2.59 & $-$2.43 & C\\
 & 5369.96 & 4.37 & $-$0.68 & L &  & 6380.74 & 4.19 & $-$1.32 & L\\
 & 5373.70 & 4.47 & $-$0.71 & L &  & 6392.53 & 2.28 & $-$4.03 & C\\
 & 5389.47 & 4.42 & $-$0.25 & L &  & 6393.60 & 2.43 & $-$1.43 & L\\
 & 5400.50 & 4.37 & $-$0.10 & L &  & 6411.65 & 3.65 & $-$0.66 & L\\
 & 5410.91 & 4.47 & $+$0.40 & L &  & 6419.95 & 4.73 & $-$0.09 & L\\
 & 5417.03 & 4.42 & $-$1.53 & L &  & 6421.35 & 2.28 & $-$2.01 & L \\
 & 5434.52 & 1.01 & $-$2.12 & L &  & 6430.85 & 2.17 & $-$2.01 & L\\
 & 5441.33 & 4.31 & $-$1.58 & L &  & 6436.40 & 4.19 & $-$2.46 & C\\
 & 5445.04 & 4.39 & $+$0.04 & L &  & 6469.19 & 4.84 & $-$0.62 & L \\
 & 5487.75 & 4.32 & $-$0.65 & L &  & 6518.37 & 2.83 & $-$2.30 & C \\
 & 5497.52 & 1.01 & $-$2.84 & L &  & 6551.67 & 0.99 & $-$5.79 & C \\
 & 5506.78 & 0.99 & $-$2.80 & L &  & 6574.22 & 0.99 & $-$5.02 & C \\
 & 5522.44 & 4.21 & $-$1.40 & L &  & 6591.31 & 4.59 & $-$2.07 & C \\
 & 5531.98 & 4.91 & $-$1.46 & L &  & 6592.91 & 2.72 & $-$1.47 & L \\
 & 5554.90 & 4.55 & $-$0.38 & L &  & 6593.87 & 2.44 & $-$2.42 & L \\
 & 5560.21 & 4.43 & $-$1.04 & L &  & 6597.56 & 4.79 & $-$0.92 & L \\
 & 5563.60 & 4.19 & $-$0.84 & L &  & 6608.02 & 2.28 & $-$4.03 & C \\
 & 5567.39 & 2.61 & $-$2.56 & L &  & 6609.11 & 2.56 & $-$2.69 & L \\
 & 5569.62 & 3.42 & $-$0.49 & L &  & 6646.93 & 2.61 & $-$3.99 & C\\
 & 5576.09 & 3.43 & $-$0.85 & L &  & 6653.85 & 4.14 & $-$2.52 & C\\
 & 5584.77 & 3.57 & $-$2.17 & L &  & 6699.14 & 4.59 & $-$2.19 & C\\
 & 5624.02 & 4.39 & $-$1.33 & L &  & 6703.56 & 2.76 & $-$3.16 & C\\
 & 5633.94 & 4.99 & $-$0.12 & L &  & 6704.48 & 4.22 & $-$2.66 & C\\
 & 5635.82 & 4.26 & $-$1.74 & L &  & 6710.32 & 1.80 & $-$4.88 & C\\
 & 5638.26 & 4.22 & $-$0.72 & L &  & 6713.74 & 4.79 & $-$1.60 & C\\
 & 5686.53 & 4.55 & $-$0.45 & L &  & 6739.52 & 1.56 & $-$4.95 & C\\
 & 5691.50 & 4.30 & $-$1.37 & L &  & 6745.96 & 4.07 & $-$2.77 & C\\
 & 5705.46 & 4.30 & $-$1.36 & L &  & 6750.15 & 2.42 & $-$2.62 & L\\
 & 5717.83 & 4.28 & $-$0.97 & L &  & 6752.71 & 4.64 & $-$1.20 & L\\
 & 5731.76 & 4.26 & $-$1.15 & L &  & 6783.70 & 2.59 & $-$3.98 & C\\
 & 5762.99 & 4.21 & $-$0.41 & L &  & 6793.25 & 4.07 & $-$2.47 & C\\
 & 5806.73 & 4.61 & $-$0.90 & L &  & 6806.84 & 2.73 & $-$3.21 & C\\
 & 5814.81 & 4.28 & $-$1.82 & L &  & 6810.26 & 4.61 & $-$1.20 & L\\
 & 5852.22 & 4.55 & $-$1.18 & L &  & 6820.37 & 4.64 & $-$1.17 & L\\
 & 5883.81 & 3.96 & $-$1.21 & L &  & 6841.34 & 4.61 & $-$0.67 & L\\
 & 5916.24 & 2.45 & $-$2.99 & L &  & 6851.64 & 1.61 & $-$5.32 & C\\
 & 5934.65 & 3.93 & $-$1.02 & L &  & 6858.15 & 4.61 & $-$0.93 & L\\
 & 6024.06 & 4.55 & $-$0.06 & L &  Fe\,{\sc ii} & 4993.35 & 2.81 & $-$3.67 & L\\
 & 6027.05 & 4.08 & $-$1.09 & L &  & 5132.65 & 2.81 & $-$4.00 & L\\
 & 6056.01 & 4.73 & $-$0.40 & L &  & 5234.62 & 3.22 & $-$2.24 & L\\
 & 6065.48 & 2.61 & $-$1.53 & L &  & 5284.10 & 2.89 & $-$3.01 & L\\
 & 6079.00 & 4.65 & $-$0.97 & L &  & 5325.56 & 3.22 & $-$3.17 & L\\
 & 6082.71 & 2.22 & $-$3.58 & L &  & 5414.04 & 3.22 & $-$3.62 & L\\
 & 6093.64 & 4.61 & $-$1.35 & L &  & 5425.25 & 3.20 & $-$3.21 & L\\
 & 6096.66 & 3.98 & $-$1.78 & L &  & 5534.83 & 3.25 & $-$2.77 & L\\
 & 6120.24 & 0.91 & $-$5.95 & L &  & 5991.37 & 3.15 & $-$3.56 & L\\
 & 6136.62 & 2.45 & $-$1.40 & L &  & 6084.09 & 3.20 & $-$3.80 & L\\
 & 6137.69 & 2.59 & $-$1.40 & L &  & 6149.25 & 3.89 & $-$2.72 & L\\
 & 6137.69 & 2.59 & $-$1.40 & L &  & 6247.55 & 3.89 & $-$2.34 & L\\
 & 6151.62 & 2.18 & $-$3.29 & L &  & 6416.92 & 3.89 & $-$2.68 & L\\
 & 6157.73 & 4.08 & $-$1.11 & L &  & 6432.68 & 2.89 & $-$3.58 & L\\
 & 6165.36 & 4.14 & $-$1.47 & L &  &         &      &         &\\\hline
\end{tabular}
\par C:Castro et al. (1997);\,L:Lambert et al. (1996)
\end{table*}

\begin{table*}
\caption{Atmospheric parameters of the studied stars. Source is the same as in Table 1.}
\begin{tabular}{ccccccc}\hline
Star & $T_{\rm eff}$ (K) & $\log g$ & [Fe\,{\sc i}/H]$\pm$ $\sigma$(\#)  & 
[Fe\,{\sc ii}/H]$\pm$ $\sigma$(\#) & $\xi$(km\,s$^{-1}$)  & Source \\\hline
BD-08$^\circ$3194 & 4900 & 3.0 & $-$0.10$\pm$0.16(36) & $-$0.05$\pm$0.21(13)& 1.6 & I\\
BD-09$^\circ$4337 & 4800 & 2.6 & $-$0.24$\pm$0.21(28) & $-$0.27$\pm$0.21(7)  & 2.7 &\\
BD-14$^\circ$2678 & 5200 & 3.1 & $+$0.01$\pm$0.12(43) & $+$0.01$\pm$0.11(11) & 1.4 &\\
CD-27$^\circ$2233 & 4700 & 2.4 & $-$0.25$\pm$0.18(51) & $-$0.24$\pm$0.17(11) & 1.4 &\\
CD-29$^\circ$8822 & 5100 & 2.8 & $+$0.04$\pm$0.15(61) & $+$0.03$\pm$0.07(10) & 1.3 &\\
CD-30$^\circ$8774 & 4900 & 2.3 & $-$0.11$\pm$0.14(43) & $-$0.11$\pm$0.09(11) & 1.2 &\\
CD-38$^\circ$585  & 4800 & 2.6 & $-$0.52$\pm$0.09(59) & $-$0.57$\pm$0.09(12) & 1.2 &\\
CD-42$^\circ$2048 & 4400 & 1.6 & $-$0.23$\pm$0.16(31) & $-$0.23$\pm$0.18(8)  & 1.6 &\\
CD-53$^\circ$8144 & 4800 & 2.3 & $-$0.19$\pm$0.15(54) & $-$0.16$\pm$0.17(9)  & 1.6 &\\
CD-61$^\circ$1941 & 4800 & 2.4 & $-$0.20$\pm$0.14(73) & $-$0.17$\pm$0.10(12) & 1.3 &\\\hline
\end{tabular}

\par Table 3 is published in its entirety in the electronic edition 
of the Monthly Notices of the Royal Astronomical Society.
A portion is shown here for guidance regarding its form and content.
\end{table*}

\begin{table*} 
\caption{Mean temperature, surface gravity, metallicity and microturbulent velocity and standard 
deviations ($\sigma$) for the barium giant stars analyzed in this work and the field giants 
previously analyzed. We show the values based on the mean from the gaussian fits
(labelled as ``g'') and the mean values based on the analysis of the total sample 
(labelled as ``m'').}
\begin{tabular}{cccccc}\hline                   
Sample & $\langle$$T_{\rm eff}$$\rangle$,K & $\langle$$\log g$$\rangle$ & $\langle$[Fe/H]$\rangle$ & $\langle$$\xi$$\rangle$,km\,s$^{-1}$ & Mean\\
                & $\sigma$ & $\sigma$    & $\sigma$  & $\sigma$ & \\\hline
Barium giants   &  4\,947/4\,428  &  2.45/1.60/1.17  &  $-$0.12/$-$0.49 &  1.44 & g \\
                &   145/146       &  0.31/0.11/0.07  & \,\,0.14/0.09    &  0.15 &   \\\hline

Field giants    &  4950       &  2.74       &  $-$0.02  &  1.52 & g \\
                &   176       &  0.33       & \,\,0.15  &  0.24 &   \\\hline\hline

Sample & $\langle$$T_{\rm eff}$$\rangle$,K & $\langle$$\log g$$\rangle$ & $\langle$[Fe/H]$\rangle$ 
& $\langle$$\xi$$\rangle$ km\,s$^{-1}$\\
                & $\sigma$ & $\sigma$    & $\sigma$  & $\sigma$ & \\\hline
Barium giants   &  4\,800 &  2.30  &  $-$0.19 &  1.45 & m \\
                &   260   &  0.48  & \,\,0.24 &  0.27 & \\\hline

Field giants    &  4840   &  2.62  &  $-$0.10  &  1.59 & m\\
                &   263   &  0.40  & \,\,0.18  &  0.38 & \\\hline

\end{tabular}
\end{table*}

\begin{table*} 
\caption{Atmospheric parameters from the literature.}
\begin{tabular}{cccccc}\hline
Star & $T_{\rm eff}$ (K) & $\log g$ & [Fe\,{\sc i}/H]   & $\xi$(km\,s$^{-1}$)  & Source \\\hline

 BD+09$^\circ$2384   & 4900 &  2.5 & $-$0.98  &  1.2 &  (1) \\
                    & 5200 &  3.0 & $-$0.71  &  2.0 &  (2) \\\hline

 CPD-64$^\circ$4333  & 4900 &  2.6 & $-$0.10  &  1.4  & (1) \\
                    & 4800 &  2.4 & $+$0.05  &       & (3) \\\hline

 HD 749           & 4700 &  2.6 & $-$0.29  &  1.3 &  (1) \\
                  & 4580 &  2.3 & $-$0.06  &  0.9 &  (4) \\\hline

 HD 4084          & 4800 &  2.8 & $-$0.42  &  2.2 &  (1) \\
                  & 4800 &  2.0 & $-$0.70  &  2.5 &  (5) \\\hline

 HD 5424          & 4700 &  2.4 & $-$0.41  &   1.1  & (1) \\
                  & 4700 &  1.8 & $-$0.51  &   1.1  & (4) \\
                  & 4600 &  2.3 & $-$0.21  &        & (6) \\\hline
\end{tabular}

\par Table 5 is published in its entirety in the electronic edition 
of the Monthly Notices of the Royal Astronomical Society.
A portion is shown here for guidance regarding its form and content.
\end{table*}

\begin{figure} 
\includegraphics[width=\columnwidth]{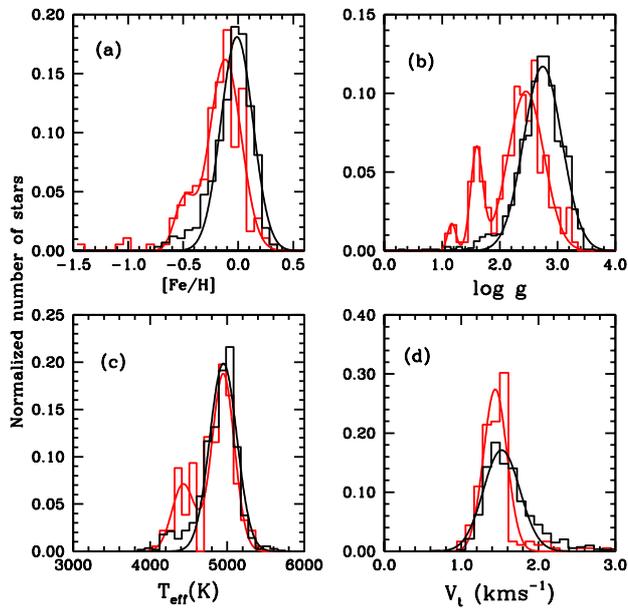}
\caption{Normalized histograms for the barium stars (red) and 
field giants of the literature (black). For the metallicity (a), we also show the 
two fits for the gaussian distribution for the barium and one single gaussian 
fit for this sample of field giants. For the surface gravity (b), there are 
three gaussians fitting three peaks of the distribution for the barium stars. 
For the temperature (c), we show two peaks of the distribution
for the barium stars. In (d) we show one gaussian fit for microturbulent velocity.}
\end{figure}

\begin{figure} 
\includegraphics[width=\columnwidth]{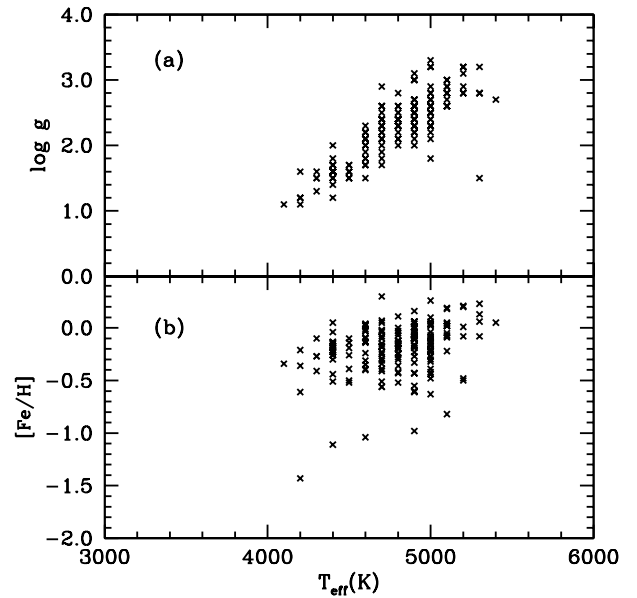}
\caption{Relations between $\log g$ {\sl versus} $T_{\rm eff}$ (a) and 
[Fe/H] {\sl versus} $T_{\rm eff}$ (b) for the barium stars analyzed in this work.}
\end{figure}

\begin{figure} 
\includegraphics[width=\columnwidth]{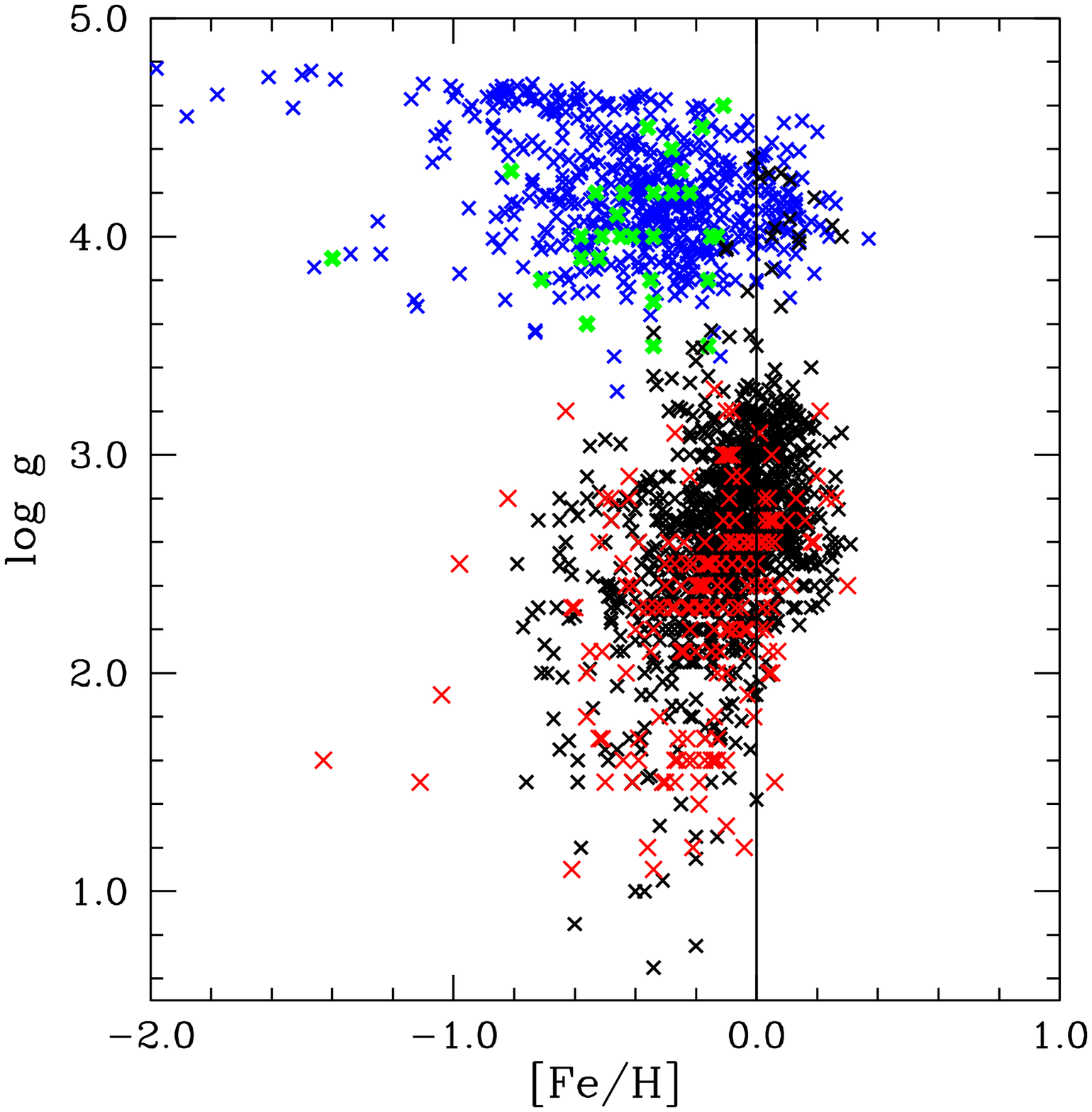}
\caption{$\log g$ {\sl versus} metallicity ($[$Fe/H$]$) diagram for the same 
field giants of Figure 3 (black crosses), barium stars (red crosses), field dwarf 
(blue crosses) and barium dwarfs (green crosses). The red cross at 
$\log g$\,=\,1.5 with [Fe/H]\,=\,0.04 is the star HD 204075.}  
\end{figure}

\subsection{Search for technetium lines in barium stars}

\par The stars with gravities between 1.1 and 1.7 are the most evolved
barium stars of our sample and will probably become extrinsic S
stars, that are considered the cooler descendents of
barium stars (Jorissen et al. 1998). The extrinsic S stars are binaries
and Tc-poor stars, contrary to the single, high luminosity and
Tc-self-enriched AGB S stars (Jorissen et al. 1993). Motivated by this
S star dichotomy, we investigated whether these evolved barium stars
of our sample show any indication of the presence of a technetium line in their spectra.
Although we have shown that all the
barium stars analyzed in this work are not luminous enough, like
AGB stars, to be self-enriched in the elements created by the s-process
(Section 5.1), the non-detection of technetium lines would also give
support to the binary transfer hypothesis. Among the 35 evolved
barium stars, we searched for the technetium lines at 4238.19\,\AA\,,
4262.27\,\AA\, and 5924.47\,\AA\,. We used some
published spectra of S stars where these lines have been unambiguously
detected (Sanner 1978; Smith \& Lambert 1986; Van Eck \& Jorissen
1999), as well as The Solar Spectrum (Moore et al. 1966).

\par In Figures 6, 7 and 8, we show the spectra of two representative
evolved barium stars, HD 43389 and HD 66291, in the regions around the
three technetium lines mentioned above. The location of Tc\,{\sc i}
transitions are shown as solid lines.  In Figure 6, it is clear that
the technetium line at 4238.19\,\AA\, is absent in the spectra of both
HD 43389 and HD 66291.  In Figure 7, we can identify a feature at
4262.24\,\AA\,. However, following the same arguments raised by Little
et al. (1987) about the likelihood of the presence of Tc\,{\sc i}, we
may conclude that this feature is ``too weak and too badly blended to
be identified with Tc''.  In addition, as raised by Van Eck \&
Jorissen (1999), there is a Nd\,{\sc ii} line at 4262.228\AA\,
(indicated by as solid line just before the technetium transition in
Figure 7) that is probably present in this feature, since these stars
are s-process enriched. Finally, Figure 8 does not show any sign of
the presence of the technetium transition at 5924.47\,\AA\,.

\par Other authors have searched for technetium in barium stars. In particular,
Little-Marenin \& Little (1987) searched for Tc\,{\sc ii} lines in the ultraviolet in the spectrum
of HD 116713 (=\,HR 5058), but they did not find any of their candidate
lines. Smith (1984) did not detect the technetium line at
5924.47\,\AA\, among his sample of stars either. Smith \& Wallerstein (1983)
also failed to detect this line in the spectrum of the evolved barium
star HD 178717. The results presented here indicate that none of the evolved barium stars 
of our sample showed the presence of any technetium line in their spectra, in agreement with 
previous findings. 

\begin{figure} 
\includegraphics[width=\columnwidth]{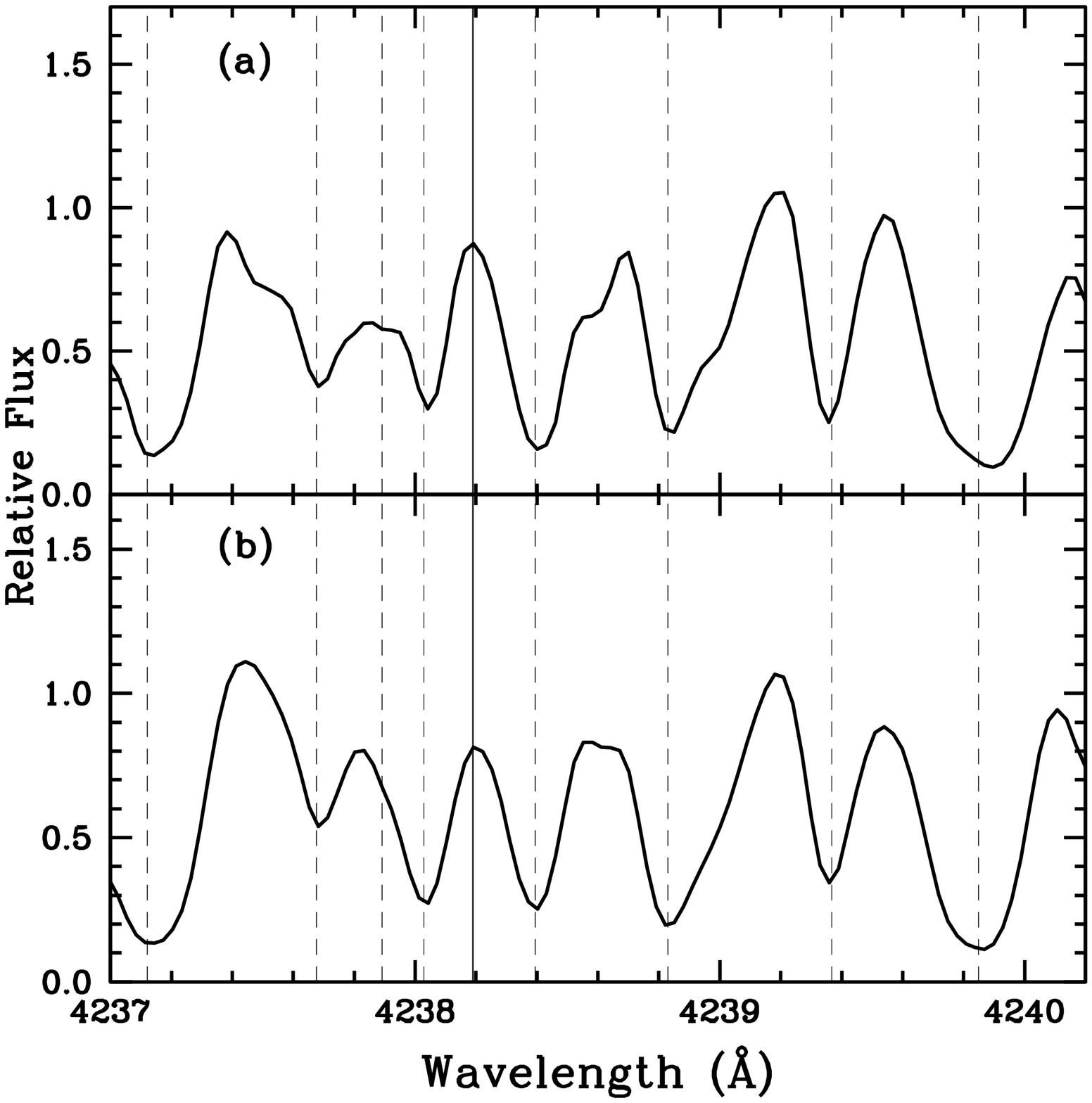}
\caption{Spectra of HD 43389 (a) and HD 66291 (b).
Absorption lines due to the transitions of Fe\,{\sc i} 4237.12,
Fe\,{\sc i} 4237.68, Ti\,{\sc i} 4237.89, Fe\,{\sc i} 4238.03, 
La\,{\sc ii}$+$CH 4238.39, Fe\,{\sc i} 4238.83, Fe\,{\sc i} 4239.37 
and Fe\,{\sc i} 4239.85 are shown. Dashed lines represent 
their rest wavelengths. We also show the technetium transition at a rest 
wavelength of 4238.19 (solid line).}
\end{figure}

\begin{figure} 
\includegraphics[width=\columnwidth]{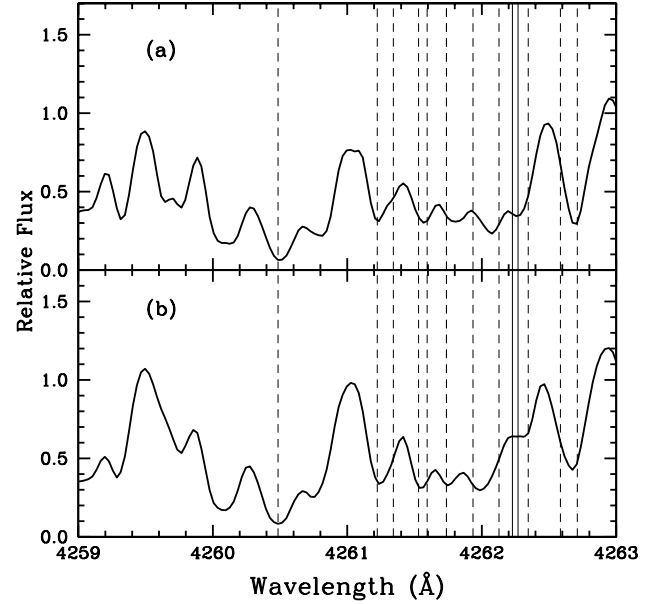}
\caption{Spectra of HD 43389 (a) and HD 66291 (b).
Absorption lines due to the transitions of Fe\,{\sc i} 4260.49,
CH 4261.22, Cr\,{\sc i} 4261.34, CH 4261.53, Ti\,{\sc i} 4261.59,
CH 4261.74, Cr\,{\sc ii}$+$CH 4261.94, Cr\,{\sc i}$+$Gd\,{\sc ii} 4262.13,
Cr\,{\sc i} 4262.35, and the CH lines at 4262.59 and 4262.71 are also shown.
Dashed lines represent their rest wavelengths. We also show
a line of technetium at a rest wavelength of 4262.27 and a neighbour 
Nd\,{\sc ii} line at 4262.228\AA\, (solid lines).}
\end{figure}

\begin{figure} 
\includegraphics[width=\columnwidth]{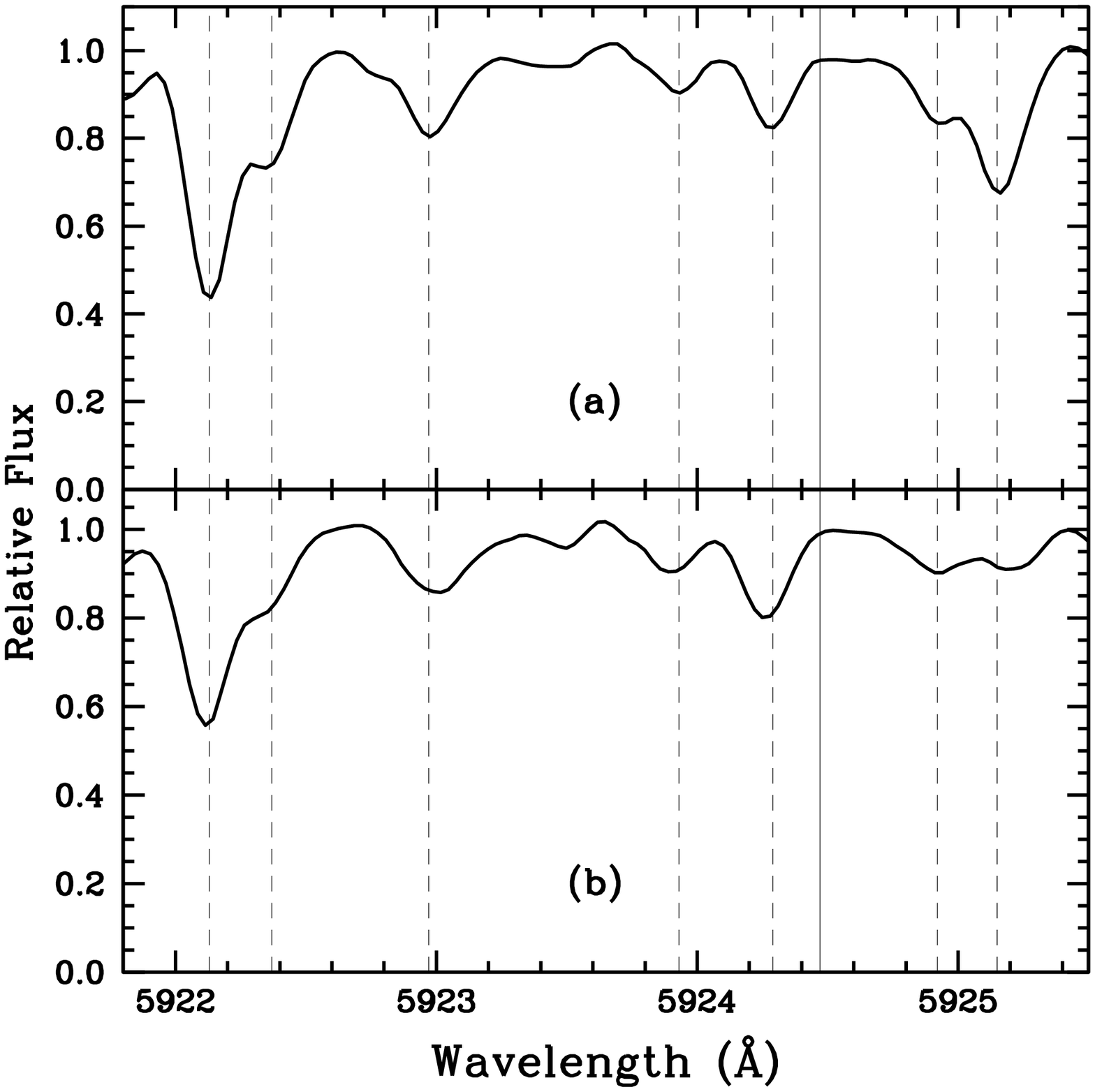}
\caption{Spectra of HD 43389 (a) and HD 66291 (b).
Absorption lines due to the transitions of Ti\,{\sc i} 5922.12 ,
CN$+$Co\,{\sc i} 5922.37, CN$+$ Ce\,{\sc ii} 5922.95, Ni\,{\sc i} 5923.93.
CN 5924.29, Ce\,{\sc ii}$+$CN 5924.92, and Zr\,{\sc ii} 5925.13
are shown. Dashed lines represent their rest wavelengths. 
We also show a transition of technetium at a rest wavelength of 5924.47 
(solid line).}
\end{figure}

\subsection {Abundance analysis}                                        

\par The abundances of both the barium and the rejected barium stars
were determined in the same way as in our previous studies (Pereira et
al. 2011). The equivalent widths were calculated by integration
through a model atmosphere and were compared to the observed
equivalent widths. The calculation is repeated, changing the abundance
of the element in question, until a match is achieved. We used the
current version of the line-synthesis code {\sc moog} (Sneden 1973) to
carry on the calculations. The line lists were also the same used in
our previous studies of barium giants.  The adopted abundances for the
elements analyzed in this work were normalized to the Solar abundances
of Grevesse \& Sauval (1998). For the solar iron abundance, we adopted
$\log \varepsilon$(Fe)\,=\,7.52. Table 6 shows the atomic lines used
to derive the abundances of the elements. We also provide the
reference for the $\log gf$ values used in the abundance
determination. For the radiative and Stark broadening, we used the
standard option available in {\sc MOOG}, while for the collisional
broadening we used the Uns\"old approximation for all the lines.
Tables~7 and 8 provide the derived abundances, in the notation [X/Fe],
for both the barium and the rejected barium stars.  The seventh column
in Table 8 gives the {\sl mean} abundance ratio of the s-process
elements ([Y/Fe], [Zr/Fe], [La/Fe], [Ce/Fe], and [Nd/Fe]) in the
notation [s/Fe], where ``s'' represents the mean abundance of the
element synthesized by the s-process. The eighth column in Table 8
gives the {\sl mean} [hs/ls] ratio, i.e. the mean abundance ratio of
the heavier elements of the s-process ([La/Fe], [Ce/Fe], and [Nd/Fe])
{\sl minus} the mean abundance ratio of the lighter elements of the
s-process ([Y/Fe] and [Zr/Fe]).

\par Only spectral lines with equivalent widths between
10\,m\AA\, and 150\,m\AA\, were used in the abundance determination.
We did not measure the barium abundance in our stars because all barium
lines are very strong and have equivalent widths between 200\,-\,400
m\AA\, and in some cases even larger than that ($\sim$1.0\,\AA).
Therefore, these barium lines are not at the linear part of the curve of
growth (Hill et al. 1995; Pereira et al, 2011, Katime Santrich et
al. 2013).  In addition, these very strong barium lines are quite
sensitive to the microturbulence, and they may also be affected by non-LTE effects
(Antipova et al. 2003).  However, since we have measured several 
lines of other elements synthesized by the s-process (Y, Zr, La, Ce
and Nd), we believe that we have probed this nucleosynthesis process fairly
well.

\begin{table*} 
\caption{Absorption lines used for abundance determination.}
\begin{tabular}{ccccc|ccccc}\hline
Ion   &  $\lambda$(\AA)  &  $\chi$(eV) & log$gf$   & Ref. & Ion &  $\lambda$(\AA) & $\chi$(eV) & log$gf$ & Ref.
\\\hline
Na\,{\sc i} & 5682.65 & 2.10 & $-$0.70 &  PS  &  Ca\,{\sc i} & 5581.79 & 2.52 & $-$0.67 &C2003\\
Na\,{\sc i} & 5688.22 & 2.10 & $-$0.40 &  PS  &  Ca\,{\sc i} & 5601.28 & 2.52 & $-$0.52 &C2003\\
Na\,{\sc i} & 6154.22 & 2.10 & $-$1.51 &  R03 &  Ca\,{\sc i} & 5857.45 & 2.93 & $+$0.11 &C2003\\
Na\,{\sc i} & 6160.75 & 2.10 & $-$1.21 &  R03 &  Ca\,{\sc i} & 5867.57 & 2.93 & $-$1.61 &C2003\\
Mg\,{\sc i} & 4730.04 & 4.34 & $-$2.39 & R03  &  Ca\,{\sc i} & 6102.72 & 1.88 & $-$0.79 &D2002\\
Mg\,{\sc i} & 5711.10 & 4.34 & $-$1.75 & R99  &  Ca\,{\sc i} & 6122.23 & 1.89 & $-$0.32 &D2002\\
Mg\,{\sc i} & 6318.71 & 5.11 & $-$1.94 & Ca07 &  Ca\,{\sc i} & 6161.30 & 2.52 & $-$1.27 &E93  \\
Mg\,{\sc i} & 6319.24 & 5.11 & $-$2.16 & Ca07 &  Ca\,{\sc i} & 6162.18 & 1.90 & $-$0.09 &D2002\\
Mg\,{\sc i} & 6319.49 & 5.11 & $-$2.67 & Ca07 &  Ca\,{\sc i} & 6166.44 & 2.52 & $-$1.14 &R03  \\
Mg\,{\sc i} & 6765.45 & 5.75 & $-$1.94 & MR94 &  Ca\,{\sc i} & 6161.30 & 2.52 & $-$1.27 & E93 \\\hline
\end{tabular}

\par Table 6 is published in its entirety in the electronic edition 
of the Monthly Notices of the Royal Astronomical Society.
A portion is shown here for guidance regarding its form and content.
\end{table*}

\begin{table*}
\caption{Abundance ratios [X/Fe] for the  elements from Na to Ni. Source is the 
same as in Table 1.}
\begin{tabular}{|l|c|c|c|c|c|c|c|c|c|}\hline
Star & [Na/Fe] & [Mg/Fe] & [Al/Fe] & [Si/Fe] & [Ca/Fe] & [Ti/Fe] & [Cr/Fe] &  [Ni/Fe] & Source
\\\hline
BD-08$^\circ$3194  &  0.12  &  0.15   & $-$0.01 &  0.09  & 0.05  &  0.09  &  0.12  & 0.03 & I\\
BD-09$^\circ$4337  &  0.42  & $-$0.10 &  0.32   &  0.00  & 0.21  &  0.32  &  ---   & 0.09 &\\
BD-14$^\circ$2678  &  0.01  &  0.01   &  0.09   &  0.10  & 0.07  &  0.01  &  0.00  & 0.00 &\\
CD-27$^\circ$2233  &  0.19  &  0.16   &  0.10   &  0.28  & 0.15  &  0.12  & $-$0.03& 0.11 &\\
CD-29$^\circ$8822  &  0.10  &  0.02   &  0.09   &  0.13  & 0.09  & $-$0.09& $-$0.02& 0.01 &\\
CD-30$^\circ$8774  &  0.24  &  0.10   &  0.08   &  0.23  & 0.14  &  0.00  &  0.00  & 0.04 &\\
CD-38$^\circ$585   &  0.21  &  0.31   &  0.39   &  0.33  & 0.27  &  0.11  &  0.04  & 0.09 &\\
CD-42$^\circ$2048  &  0.44  &  0.28   &  0.37   &  0.36  & 0.26  &  0.18  &  ---   & 0.05 &\\
CD-53$^\circ$8144  &  0.16  &  0.17   &  0.17   &  0.11  & 0.08  &  0.08  & $-$0.09& 0.06 &\\
CD-61$^\circ$1941  &  0.06  &  0.08   &  0.08   &  0.18  & 0.04  &  0.01  & $-$0.02& 0.01 &\\\hline
\end{tabular}

\par Table 7 is published in its entirety in the electronic edition 
of the Monthly Notices of the Royal Astronomical Society.
A portion is shown here for guidance regarding its form and content.
\end{table*}

\begin{table*}
\caption{Abundance ratios [X/Fe] for the s-process elements and their mean abundance 
of the s-process and the $[$hs/ls$]$ ratio. Source is the same as in Table 1.}
\begin{tabular}{|l|c|c|c|c|c|c|c|c|}\hline
star                & [Y/Fe] & [Zr/Fe]  & [La/Fe] & [Ce/Fe] & [Nd/Fe] & [s/Fe] & [hs/ls]  & Source\\\hline
BD-08$^\circ$3194   &  0.95  &  0.95    &  1.94   &  1.27   &  1.35   &   1.29  &   0.57  &I\\
BD-09$^\circ$4337   &  1.11  &  1.51    &  ---    &  1.44   &  1.56   &   1.41  &   0.19  &\\
BD-14$^\circ$2678   &  1.02  &  0.85    &  1.08   &  0.87   &  0.87   &   0.94  &   0.01  &\\
CD-27$^\circ$2233   &  0.89  &  0.73    &  1.41   &  0.94   &  0.96   &   0.99  &   0.29  &\\
CD-29$^\circ$8822   &  1.06  &  0.81    &  1.49   &  1.06   &  0.91   &   1.07  &   0.22  &\\
CD-30$^\circ$8774   &  0.73  &  0.27    &  0.59   &  0.38   &  0.31   &   0.46  &$-$0.07  &\\
CD-38$^\circ$585    &  1.05  &  0.95    &  1.66   &  1.42   &  1.55   &   1.33  &   0.54  &\\
CD-42$^\circ$2048   &  0.95  &  0.96    &  1.56   &  1.12   &  1.22   &   1.16  &   0.34  &\\
CD-53$^\circ$8144   &  0.83  &  0.80    &  1.51   &  0.97   &  1.00   &   1.02  &   0.34  &\\
CD-61$^\circ$1941   &  0.73  &  0.68    &  1.56   &  1.15   &  1.08   &   1.04  &   0.56  &\\\hline
\end{tabular}

\par Table 8 is published in its entirety in the electronic edition 
of the Monthly Notices of the Royal Astronomical Society.
A portion is shown here for guidance regarding its form and content.
\end{table*}

\subsection{Abundances uncertainties}    

\par The abundances uncertainties for three stars, BD-14$^\circ$2678,
HD 119185, and HD 130255, are summarized in Cols.~2 to 6 of Tables
9\,-\,11. These three stars have very different atmospheric
parameters, but the uncertainties in the abundances of the elements
are similar.  The uncertainties due to the errors in each stellar
atmospheric parameter, $T_{\rm eff}$, $\log$~g, $\xi$, and [Fe/H]
(Cols.~2 to 5), were estimated by changing these parameters one at a
time by their standard errors and then computing the changes incurred
in the elements abundances. This technique has been applied to the
abundances determined from equivalent widths as well as to the
abundances determined via spectral synthesis. The abundances
uncertainties due to the errors in the equivalent width measurements
(Col.~6) were computed from the expression given by Cayrel (1988). The
errors in the equivalent widths were basically set by the S/N ratio
and the resolution of the spectra. In our case, having R\,=\,48\,000
and typical S/N ratio of 100, the expected uncertainties in the
equivalent widths are about 2--3 m{\AA}. For all the measured
equivalent widths, these uncertainties led to errors in the abundances
which are smaller than those derived from the uncertainties in the
stellar parameters.  The final uncertainties of the abundances
(Col.~7) were calculated as the root squared sum of the individual
uncertainties due to the errors in each atmospheric parameter and in
the equivalent width, under the assumption that these individual
uncertainties are independent.

\par The last column of Tables 9\,-\,11 shows the observed abundance
dispersion among the lines for those elements with more than three
available lines.  The mean value of the differences between the
calculated and the observed uncertainties (calculated {\sl minus}
observed) are not larger than $\sim 0.2$ dex (for chromium in HD
119185). For BD-14$^\circ$2678, HD 119185 and HD 130255, the mean
differences are, respectively, 0.04$\pm$0.09, 0.07$\pm$0.06, and
0.07$\pm$0.06. This indicates a good agreement between these two error
estimates.

\par From the results shown in these Tables, we verify the well-known
relations that neutral elements are more sensitive to the temperature
variations, while singly-ionized elements are more sensitive to the
$\log g$ variations.  For the elements whose abundance is based on
stronger lines, such as the lines of calcium, chromium, nickel and
yttrium, and sometimes cerium and neodymium, the error introduced by
the microturbulence is significant.

\begin{table*}
\caption{Abundance uncertainties for BD-14$^\circ$2678 which have 
$T_{\rm eff}$\,=\,5\,200\,K, $\log g$\,=\,3.1, [Fe\,{\sc i}/H]\,=\,0.01
and $\xi$\,=\,1.4 km\,s$^{-1}$. The second
column gives the variation of the abundances caused by the variation in
$T_{\rm eff}$. The other columns refer to the variations in the abundances caused by
 variations in $\log g$, $\xi$, [Fe/H], and $W_\lambda$. The seventh column gives
the compounded rms uncertainty of the second to sixth column. The last column gives the 
abundances dispersion observed among the lines for those elements with more than 
three available lines.}
\begin{tabular}{lccccccc}\hline\hline
Species & $\Delta T_{\rm eff}$ & $\Delta\log g$ & $\Delta\xi$ & $\Delta$[Fe/H] & 
$\Delta W_{\lambda}$ & $\left( \sum \sigma^2 \right)^{1/2}$ & $\sigma_{\rm obs}$\\
$_{\rule{0pt}{8pt}}$ & $+100$~K & $+0.2$ & $+$0.3 km\,s$^{-1}$ & $+$0.1 & $+$3 m\AA &  \\
\hline     
Fe\,{\sc i}    & $+$0.08  &    0.00 & $-$0.12 &    0.00 & $+$0.06 & 0.16 & 0.11 \\
Fe\,{\sc ii}   & $-$0.04  & $+$0.11 & $-$0.10 & $+$0.05 & $+$0.07 & 0.17 & 0.10 \\
Na\,{\sc i}    & $+$0.06  & $-$0.01 & $-$0.05 & $-$0.01 & $+$0.04 & 0.10 & ---  \\
Mg\,{\sc i}    & $+$0.04  & $-$0.01 & $-$0.05 &    0.00 & $+$0.04 & 0.08 & 0.18 \\
Al\,{\sc i}    & $+$0.05  & $-$0.02 & $-$0.05 & $-$0.01 & $+$0.04 & 0.09 & 0.24 \\
Si\,{\sc i}    &    0.00  & $+$0.02 & $-$0.05 & $+$0.01 & $+$0.05 & 0.07 & 0.09 \\
Ca\,{\sc i}    & $+$0.09  & $-$0.02 & $-$0.11 & $-$0.00 & $+$0.06 & 0.15 & 0.12 \\
Ti\,{\sc i}    & $+$0.13  &    0.00 & $-$0.09 & $-$0.01 & $+$0.06 & 0.17 & 0.15 \\
Cr\,{\sc i}    & $+$0.13  & $-$0.02 & $-$0.18 & $-$0.02 & $+$0.06 & 0.23 & 0.05 \\
Ni\,{\sc i}    & $+$0.06  & $+$0.02 & $-$0.08 & $+$0.01 & $+$0.06 & 0.12 & 0.07 \\
Y\,{\sc ii}    & $+$0.01  & $+$0.07 & $-$0.20 & $+$0.04 & $+$0.08 & 0.23 & 0.09 \\
Zr\,{\sc i}    & $+$0.14  & $-$0.01 & $-$0.03 & $-$0.01 & $+$0.07 & 0.16 & 0.15 \\
La\,{\sc ii}   & $+$0.01  & $+$0.09 & $-$0.12 & $+$0.04 & $+$0.07 & 0.17 & 0.07 \\
Ce\,{\sc ii}   & $+$0.01  & $+$0.08 & $-$0.16 & $+$0.04 & $+$0.08 & 0.20 & 0.17 \\
Nd\,{\sc ii}   & $+$0.03  & $+$0.08 & $-$0.21 & $+$0.03 & $+$0.04 & 0.23 & 0.15 \\\hline
\hline
\end{tabular}
\end{table*}

\begin{table*}
\caption{Same as Table 9 but for the star HD 119185, which have 
$T_{\rm eff}$\,=\,4\,800\,K, $\log g$\,=\,2.0, [Fe\,{\sc i}/H]\,=\,$-$0.43
and $\xi$\,=\, 1.3 km\,s$^{-1}$.}
\begin{tabular}{lccccccc}\hline\hline
Species & $\Delta T_{\rm eff}$ & $\Delta\log g$ & $\Delta\xi$ & $\Delta$[Fe/H] & 
$\Delta W_{\lambda}$ & $\left( \sum \sigma^2 \right)^{1/2}$ & $\sigma_{\rm obs}$\\
$_{\rule{0pt}{8pt}}$ & $+100$~K & $+0.2$ & $+$0.3 km\,s$^{-1}$ & $+$0.1 & $+$3 m\AA &  \\
\hline            
Fe\,{\sc i}    & $+$0.09  & $+$0.01 & $-$0.13 &    0.00 & $+$0.01 & 0.16 & 0.09 \\
Fe\,{\sc ii}   & $-$0.05  & $+$0.13 & $-$0.11 & $-$0.04 & $+$0.07 & 0.19 & 0.10 \\
Na\,{\sc i}    & $+$0.09  & $-$0.01 & $-$0.10 &    0.00 & $+$0.04 & 0.14 & 0.14 \\
Mg\,{\sc i}    & $+$0.05  & $-$0.01 & $-$0.08 & $-$0.01 & $+$0.04 & 0.10 & 0.15 \\
Al\,{\sc i}    & $+$0.06  & $-$0.01 & $-$0.04 &    0.00 & $+$0.05 & 0.09 & 0.09 \\
Si\,{\sc i}    &    0.00  & $+$0.04 & $-$0.04 &    0.00 & $+$0.07 & 0.09 & 0.04 \\
Ca\,{\sc i}    & $+$0.11  & $-$0.01 & $-$0.18 &    0.00 & $+$0.04 & 0.21 & 0.09 \\
Ti\,{\sc i}    & $+$0.16  & $-$0.02 & $-$0.13 &    0.00 & $+$0.04 & 0.21 & 0.12 \\
Cr\,{\sc i}    & $+$0.18  &    0.00 & $-$0.20 & $+$0.02 & $+$0.07 & 0.28 & 0.09 \\
Ni\,{\sc i}    & $+$0.07  & $+$0.03 & $-$0.10 &    0.00 & $+$0.07 & 0.14 & 0.10 \\
Y\,{\sc ii}    &    0.02  & $+$0.11 & $-$0.18 & $+$0.01 & $+$0.12 & 0.24 & 0.07 \\
Zr\,{\sc i}    & $+$0.17  & $-$0.02 & $-$0.01 &    0.00 & $+$0.08 & 0.19 & 0.12 \\
La\,{\sc ii}   & $+$0.02  & $+$0.10 & $-$0.07 &    0.00 & $+$0.10 & 0.16 & 0.06 \\
Ce\,{\sc ii}   & $+$0.03  & $+$0.09 & $-$0.16 & $+$0.01 & $+$0.12 & 0.22 & 0.21 \\
Nd\,{\sc ii}   & $+$0.02  & $+$0.10 & $-$0.09 & $+$0.04 & $+$0.08 & 0.16 & 0.11 \\\hline
\end{tabular}
\end{table*}

\begin{table*}
\caption{Same as Table 9 but for the star HD 130255 which have 
$T_{\rm eff}$\,=\,4\,400\,K, $\log g$\,=\,1.5, [Fe\,{\sc i}/H]\,=\,$-$1.11
and $\xi$\,=\,1.3 km\,s$^{-1}$.}
\centering
\begin{tabular}{lccccccc}\hline\hline

Species & $\Delta T_{\rm eff}$ & $\Delta\log g$ & $\Delta\xi$ & $\Delta$[Fe/H] & 
$\Delta W_{\lambda}$ & $\left( \sum \sigma^2 \right)^{1/2}$ & $\sigma_{\rm obs}$\\
$_{\rule{0pt}{8pt}}$ & $+90$~K & $+0.2$ & $+$0.3 km\,s$^{-1}$ & $+$0.1 & $+$3 m\AA &  \\
\hline     

Fe\,{\sc i}    & $+$0.08  & $+$0.02 & $-$0.12 & $+$0.01 & $+$0.07 & 0.16 & 0.10 \\
Fe\,{\sc ii}   & $-$0.09  & $+$0.13 & $-$0.07 & $+$0.03 & $+$0.09 & 0.20 & 0.08 \\
Na\,{\sc i}    & $+$0.08  & $-$0.01 & $-$0.05 &    0.00 & $+$0.06 & 0.11 & 0.12 \\
Mg\,{\sc i}    & $+$0.07  &    0.00 & $-$0.06 &    0.00 & $+$0.06 & 0.11 & 0.11 \\
Al\,{\sc i}    & $+$0.09  & $-$0.01 & $-$0.03 & $-$0.01 & $+$0.05 & 0.11 & ---  \\
Si\,{\sc i}    & $-$0.02  & $+$0.06 & $-$0.04 & $+$0.01 & $+$0.07 & 0.10 & 0.05 \\
Ca\,{\sc i}    & $+$0.11  & $-$0.01 & $-$0.15 & $-$0.01 & $+$0.06 & 0.20 & 0.12 \\
Ti\,{\sc i}    & $+$0.16  & $-$0.02 & $-$0.13 & $-$0.01 & $+$0.06 & 0.22 & 0.06 \\
Cr\,{\sc i}    & $+$0.11  & $-$0.01 & $-$0.17 &    0.00 & $+$0.06 & 0.21 & 0.13 \\
Ni\,{\sc i}    & $+$0.05  & $+$0.04 & $-$0.07 & $+$0.01 & $+$0.08 & 0.12 & 0.10 \\
Y\,{\sc ii}    & $+$0.02  & $+$0.10 & $-$0.16 & $-$0.04 & $+$0.10 & 0.19 & 0.18 \\
Zr\,{\sc i}    & $+$0.19  & $-$0.03 & $-$0.02 &    0.00 & $+$0.08 & 0.21 & 0.12 \\
La\,{\sc ii}   & $+$0.03  & $+$0.09 & $-$0.04 & $-$0.04 & $+$0.07 & 0.13 & 0.06 \\
Ce\,{\sc ii}   & $-$0.03  & $+$0.08 & $-$0.13 & $+$0.04 & $+$0.10 & 0.19 & 0.16 \\
Nd\,{\sc ii}   & $+$0.01  & $+$0.08 & $-$0.07 & $+$0.03 & $+$0.08 & 0.14 & 0.10 \\\hline
\end{tabular}
\end{table*}

\section{Discussion}

\subsection{The position of the stars in the 
$\log g$\,-\,log\,$T_{\rm eff}$ diagram.}

\par We estimate the stellar masses of the barium and the rejected
barium stars from their position in the $\log g$\,-\,log\,$T_{\rm
  eff}$ diagram, taking into account the results for $T_{\rm
  eff}$/$\log g$ given in Table 3.  We used the evolutionary tracks of
Fagotto et al. (1994b) for metallicities of $Z$\,=\,0.02, 0.004, and
0.008. For metallicity $Z$\,=\,0.001, we used the tracks of Schaller
et al. (1992).  For those stars with [Fe/H] in the range between
$+0.17$ and $-0.11$, we used the model tracks for $Z$\,=\,0.02, which
corresponds to [Fe/H]\,=\,$+$0.03 and takes into consideration the
{\sl mean} uncertainty of 0.14$\pm$0.03 in [Fe/H] (Table 3).  For
stars with [Fe/H] in the range between $-0.23$ and $-0.51$, we used
the model tracks for $Z$\,=\,0.008, while for stars with [Fe/H] in the
range between $-0.53$ and $-0.81$, we used the model tracks for
$Z$\,=\,0.004.  Finally, for the few stars with [Fe/H] around $-$1.0
(BD+09$^\circ$2384, HD 123396 and HD 130255), we used the model tracks
for $Z$\,=\,0.001.  In each panel of Figure 9, the different
evolutionary tracks correspond to different masses: 1.0, 1.5, 2.0,
2.5, 3.0, 4.0, 5.0, 6.0~$M_{\odot}$, and we overlap to these tracks
the positions in the $\log g$ {\sl versus} $\log T_{\rm eff}$ diagram
of the stars belonging to the same metallicity group.  We then applied
a Monte Carlo method to determine the probability for each
evolutionary trajectory to fall within the maximum and minimum values
of $\log g$ and $\log T_{\rm eff}$ of a given star.  The mass of the
trajectory with the largest probability was adopted as the mass of the
given star. The error in the mass was estimated as the mass difference
between the most probable and the second most probable trajectory.
When only one trajectory has non zero probability, we attributed an
upper limit to the error of 1.0 $M_{\odot}$.  The estimated masses are
given in Table 12.

\par Most of the stars of our sample have masses between 2.0 and
3.0\,$M_{\odot}$, which is in good agreement with the model
predictions by Han et al. (1995), who found that the masses of barium
stars should range between 1.0 and 3.0\,$M_{\odot}$.  In fact, from
the gaussian fit to the histogram in Figure 10, we find a mean value
of (2.76$\pm$0.84) $M_{\odot}$ for the mass distribution of the barium
giants.  Mennessier et al. (1997) determined the masses of barium
stars using the maximum-likelihood method (Gom\'ez et al. 1997)
applied to the L\"u catalogue (1991) and the Hipparcos astrometric
data. They found that the majority of the stars in their C (``clump
giants'') and G (``subgiant and giant stars'') groups have masses
between 1.0 and 4.5\,$M_{\odot}$. Like Mennessier et al. (1997), we
also found a smaller group of massive stars, with masses higher than
four solar masses (27 stars with 4.0\,$M_{\odot}$, 5 stars with
5.0\,$M_{\odot}$ and 3 stars with 6.0\,$M_{\odot}$).  Other studies
(Antipova et al. 2003; Smiljanic et al. 2007; Pereira et al. 2011)
also determined the masses of smaller samples of barium stars.  Taking
into account the stars that our present study has in common with those
previous works, we found a good agreement in the estimated masses,
with an uncertainty of 0.5\,-\,1.0 $M_{\odot}$. Table 13 summarizes
this comparison.  Among the set of massive barium stars, one deserves
a comment: HD 204075. Its position in the $\log g$\,-\,log\,$T_{\rm
  eff}$ diagram suggests that it is a horizontal-branch star.

\par Once we estimated the masses of the stars, we were able to
determine their spectroscopic distances.  The relation between the
distance of a star to the Sun, $r$, and its temperature, gravity,
mass, $V$ magnitude, and interstellar absorption ($A_{V}$), is given
by:

\begin{eqnarray}
\log r\: ({\rm kpc}) & = & \frac{1}{2}\left(\log \frac{M_{\star}}{M_{\odot}}
+ 0.4\left(V-A_{\rm V}+BC\right) \right. \nonumber \\
& & \left. {{\,}\atop{\,}} + 4\log T_{\rm eff} - \log g - 16.5\right).
\end{eqnarray}

\par The $B$ and $V$ colors of barium stars are affected by the
so-called Bond-Neff depression, a broad absorption feature seen in the
spectra of barium stars near $\sim$4\,000\,\AA\, and extending from
4\,000 to 4\,500\,\AA\, (Bond \& Neff 1969; Gow 1976; L\"u \& Sawyer
1979). Therefore, the usual method to determine the extinction through
the relation $A_{\rm V}\,=\,3.2\times\,E(B-V)$, using the $B-V$ colors
cannot be used. We considered that stars at $\sim 100$ pc are affected
by an interstellar extinction of $A_{\rm V}$\,=\,0.1 (B\"ohm-Vitense
et al. 2000), while for the other stars the extinction was determined
using calibrations between $A_{\rm V}$, galactic coordinates and
distances given by Chen et al. (1998). Bolometric corrections were
derived from Alonso et al. (1999), assuming $M_{\rm bol\odot}\! =\!
+4.74$ (Bessel 1998).  The results are summarized in Table 12.  This
Table provides the bolometric corrections, the interstellar
absorption, and the derived distances. The sixth and seventh columns
provide, respectively, the distance given by Hipparcos parallax (van
Leeuwen 2007) and the absolute visual magnitude.

\par For six stars, HD 43389, HD 66291, HD 74950, HD 142491, HD
168986, and HD 252117, we did not determine the interstellar
absorption, nor their distances. Since they would be at more than
1\,000 pc considering null extinction, and in addition they lie at
very low galactic latitudes, $|$b$|$$\leq$10$^\circ$, the polynomial
expression to obtain $A_{\rm V}$ given by Chen et al. (1998) can not
be used for these stars.  In Figure 11, we compare our spectroscopic
distances for the barium stars with those obtained from Hipparcos
parallaxes, where a linear correlation is evident.  We note that for
distances longer than 600 pc, the errors given by Hipparcos parallaxes
are much larger than those of the estimated spectroscopid distances.
We also note that for spectroscopic distances less than 300 pc, our
distances underestimates Hipparcos distances by up ot 30\%, while for
spectroscopic distances higher than 300 pc, our distances
overestimates the Hipparcos ones by up to 50\%. We also found that the
mean quadratic difference between the spectroscopic and the Hipparcos
distances given by $\langle$($r_{\rm spec}\,-\,r_{\rm
  Hip}$)$^{2}$$\rangle$ is smaller than the mean squared sum of the
errors in each distance given by $\langle$($\Delta\,r_{\rm
  spec}$)$^{2}$$+$ ($\Delta\,r_{\rm Hip}$)$^{2}$$\rangle$, implying
that the two distances distributions are indistinguishable.

\par The luminosities of the barium giants were obtained either by
taking into consideration their distances and interstellar reddening,
or by knowing their masses from the $\log g$\,-\,log\,$T_{\rm eff}$
diagram.  This latter was the case for HD 43389, HD 66291, HD 74950,
HD 142491, HD 168986, and HD 252117. The luminosities are listed in
the eighth column of Table 12.  We found that the mean error for the
values of log $(L/L_{\odot})$ is $\pm$0.28.  For the field giants,
luminosities are only available from Luck \& Heiter (2007) and Takeda
et al. (2008).  Figure 12a shows the normalized histograms of the
luminosities of the field giants and the barium giants.  We see that
the luminosity distribution of the field giants is narrower than that
of the barium stars.  A gaussian fit to the field giants gave a mean
log $(L/L_{\odot})$ of 1.81$\pm$0.20, while for the barium stars we
obtained a mean of 2.21$\pm$0.42.  The field giants have a smaller
standard deviation than the barium giants probably because the samples
of Luck \& Heiter (2007) and Takeda et al. (2008) contain many clump
giants compared to our sample of barium giants. In fact, the
distribution of those two samples in the $\log g$\, {\sl versus}
\,log\,$L/L_{\odot}$ diagram, shown in Figure 12b, indicates that
there is a concentration or a ``clump'' around
log\,$L/L_{\odot}$\,=\,1.7\,-\,1.9.

\par In Figure 12a, we also show two vertical lines corresponding to
two different metallicities and masses.  These represent the lower
luminosoty limits for a star to develop the first thermal pulses and
to become self-enriched in the elements of the s-process, as expected
for the AGB models from Vassiliadis \& Wood (1993).  The dashed line
gives the minimum luminosity for a star with 2.5 $M_{\odot}$ and
Z\,=\,0.008, and the solid line gives the minimum luminosity for a
star with 5.0 $M_{\odot}$ and Z\,=\,0.02.  The arrows indicate the
luminosities of the barium stars with 2.5 and 6.0 $M_{\odot}$,
respectively.  The arrow at log\,$L/L_{\odot}$\,=\,3.4 corresponds to
the three stars with 6.0 $M_{\odot}$ (HD 204075, HD 216809, and HD
221879), but the minimum luminosity for a 5.0 $M_{\odot}$ mass star to
develop the first thermal pulses is log\,$L/L_{\odot}$\,=\,4.16 (solid
line).  Therefore, none of the stars in our sample are still luminous
enough to develop the thermal pulses.  Figure 12 provides the first
determination of the observational luminosity function for the barium
giants. As mentioned earlier, barium stars are warmer than AGB
stars. Therefore, ionization equilibrium provides the spectroscopic
gravity for these stars such that it is possible to obtain
``accurate'' distances even for those that do not have accurate
parallaxes. For carbon stars and hence very cool stars, where it is
not possible to obtain the spectroscopic gravity, Guandalini \&
Cristallo (2013) also determined an observational luminosity function
which was based on the Hipparcos astrometric data and/or the
period-luminosity relation.

\par With the derived luminosities, and using the theoretical
isochrones of Fagotto et al. (1994b) for the same groups of
metallicities considered for the evolutionary tracks, we determined
the ages of the barium stars following the similar Monte Carlo
approach as for the masses.  In Figure 13, we show the isochrones from
0.06 Gyr (\,=\,60 Myr) to 10.0 Gyr given in the notation of
$\log\,t\,$ (in years).  The last column of Table 12 gives the
estimated age for each star. We found that the ages of the barium
stars have a strong concentration between 320 Myr ($\log\,t\,$
(years)\,=\,8.5) and 1.0 Gyr ($\log\,t\,$(years)\,=9.0).  Our results
are in good agreement with those obtained by Mennessier et al. (1997),
who found that the ages of barium stars should lie in the range 7.8
$\leq$ $\log\,t\,$ $\leq$ 10.0 (age in years), with a strong
concentration at $\log\,t\,$ between 8.0 and 9.0.

\par In Figure 14, we show the relation between the age and the
metallicity and the relation between the age and the mass for the
barium stars.  In the age {\sl versus} metallicity diagram, although
there is a scatter in metallicity for a given age, we found an
anti-correlation between metallicity and age as illustrated by the
linear regression to the data.  This is expected from models of
galactic chemical evolution (see Figure 1 of Chiappini et al. 1997).
In the age {\sl versus} mass diagram, the red curve shows the
polynomial fit to the data represented by the red points, where each
point gives the mean mass for a given age of the barium giants. The
black curve represents the polynomial fit to the data of the field
giants analyzed by Takeda et al. (2008), where each point also gives
the mean mass for a given age.  There is a clear trend between mass
and age, that is the higher mass, the younger the star. The
explanation for this trend is that the age of these barium and field
giants is basically the time they spent on the main-sequence, which
depends on the mass.

\par With the distances of the barium stars in hand, we were also able
to determine their distances from the Galactic plane ($z$) and thus to
obtain the height scale $h$. By definition, the height scale is
obtained from an exponentially decaying distribution which is
symmetric with respect to the Galactic plane. Unfortunately, the
distribution of our sample is not symmetric but it is centered around
$z_{0}\sim-100$ pc, because most stars in our sample lie in the
southern hemisphere. Therefore, in order to estimate the height scale
we followed two different approaches. The first approach was to
consider only our data for $z$ less $\sim$-100 pc and to fit 
an exponential law $N=N_{0}e^{-\left|z\right|/h}$, extrapolated to
$z=0$. This approach has the disadvantage that we missed information
from more than 45\% of the total sample. So, in order to use
information from all the sample, a second possible approach was to fit
all the data by a Gaussian law $N=N_{0}e^{-(z-z_{0})^{2}/2\sigma^{2}}$,
and to use the standard deviation as a proxy for $h$. Nevertheless, we
must bear in mind that this approach will tend to overestimate the
height scale. Figure 15 shows the distribution of $z$ for our sample,
together with different fits, whose parameters are listed in Table 14.
From these fits, we see that the scale height of the barium stars is
between 233 and 362 parsecs, which is similar to the value of 300 parsecs found by
Gilmore \& Reid (1983) for the scale height of the Galactic disk.

\begin{figure}
\includegraphics[width=\columnwidth]{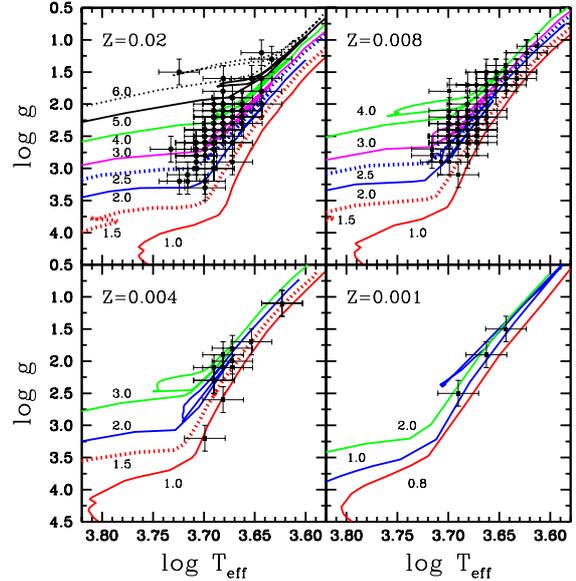}
\caption{Position of the barium stars and the rejected barium stars 
in the $\log g$\,-\,log $T_{\rm eff}$  diagram. We show the 
evolutionary tracks where the numbers correspond to stellar masses in units 
of solar mass. Models were taken from Fagotto et al. (1994b) and 
Schaller et al. (1992).}
\end{figure}

\begin{figure}
\includegraphics[width=\columnwidth]{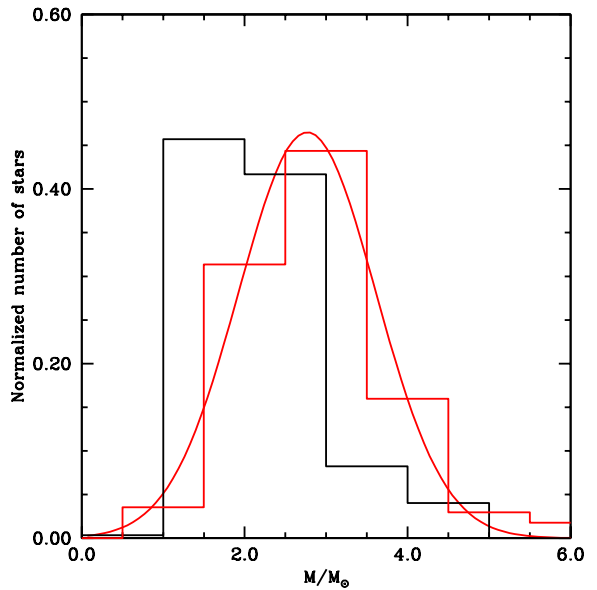}
\caption{Normalized mass histogram for the barium stars (red) and
the field giants of the literature (black). We also show the gaussian
distribution fit for the barium giants.}
\end{figure}

\begin{figure}
\includegraphics[width=\columnwidth]{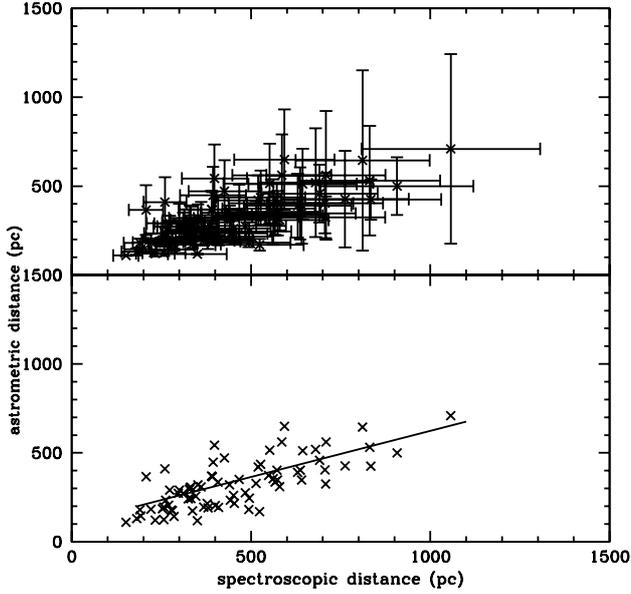}
\caption{Astrometric distance as given by Hipparcos parallax {\sl versus} 
spectroscopic distance for some of the barium stars analyzed in this work with 
error bars (upper pannel) and without error bars (lower pannel). We also show a 
linear fit between these two distance determinations.}
\end{figure} 

\begin{figure}
\includegraphics[width=\columnwidth]{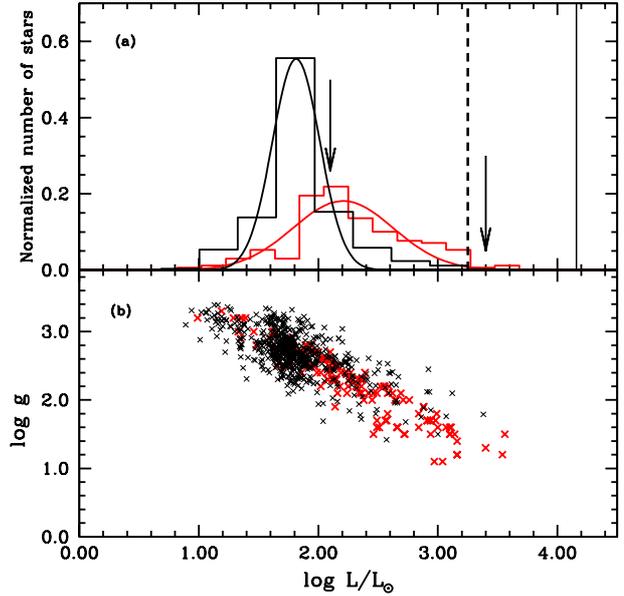}
\caption{Normalized luminosity histogram for the barium stars (red)
and the field giants of the literature (black) taken from Luck \&
Heiter (2007) and Takeda et al. (2008) (a).  We also show gaussian
distribution fits for the barium and the field giants. Vertical
lines represent the minimum luminosities of thermally pulsing AGB star
according to model predictions given by Vassiliadis \& Wood (1993). The dashed
vertical line represents the minimum luminosity for a star with the
2.5\,M$_{\odot}$ and Z\,=\,0.008 and the solid line represents the minimum
luminosity for a star of 5.0\,M$_{\odot}$ and Z\,=\,0.02. The arrows
at log $(L/L_{\odot}$)$\sim$2.1 and 3.4 show the positions of the barium 
stars with 2.5 and 6.0 M$_{\odot}$ respectively in the histogram. In (b) we show 
the position of these two samples in the $\log g$\,-\,log $(L/L_{\odot}$) diagram.}
\end{figure}

\begin{figure}
\includegraphics[width=\columnwidth]{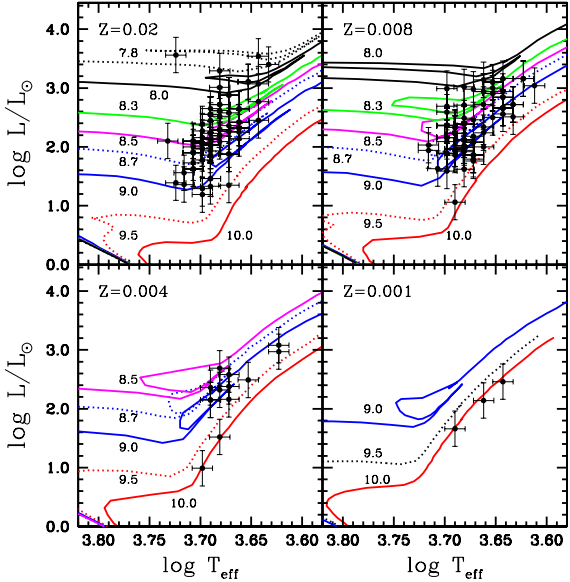}
\caption{Position of the barium stars and the rejected barium stars 
in the $\log L/L_{\odot}$\,-\,log $T_{\rm eff}$  diagram in order to obtain
their ages. Theoretical isochrones were taken from Fagotto et al. (1994b).
The numbers corresponds to stellar ages given in the notation of 
log\,t (years).}
\end{figure}

\begin{figure}
\includegraphics[width=\columnwidth]{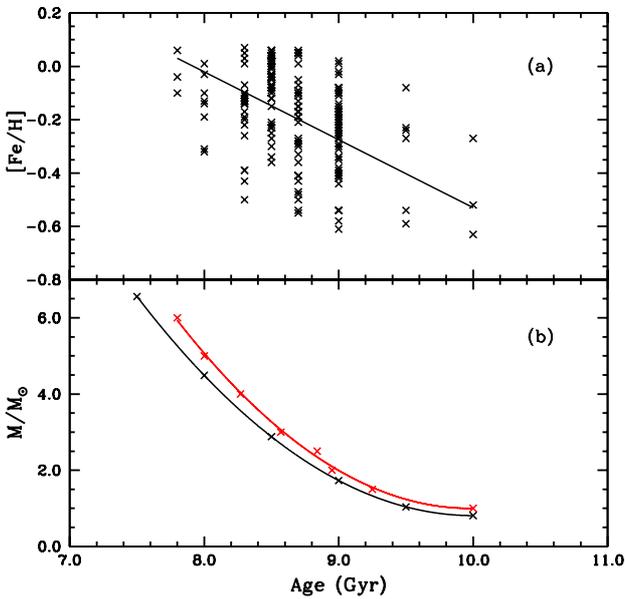}
\caption{Age\,-\,metallicity (a) and age\,-\,mass relation diagram (b) for the
barium stars analyzed in this work. In the age\,-\,mass diagram the
red points represent the mean mass for a given age fitted by
the red line.  Each data point may represents more than one star. Black
points connected by solid line represent the trend given in Takeda
et al. (2008) between mass and age.}
\end{figure} 

\begin{figure}
\includegraphics[width=\columnwidth]{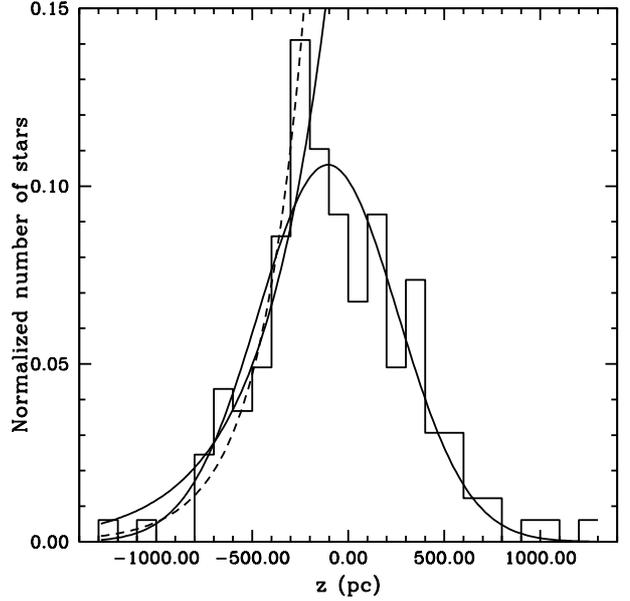}
\caption{Distribution of our barium stars sample distances to the Galactic
 plane. We note that the distribution is shifted to negative values of $z$ 
due to observational bias. Gaussian fits and exponential laws
give the best fits to the sample. For the exponential laws the lines
represent, respectively, fit to the data for stars with $z$ less than 100 parsecs
(solid line) and fit to the data for stars with $z$ less than 200 parsecs
(short dashed line). Parameters of the fits are listed in Table 14.}
\end{figure}

\begin{table*}
\caption{Stellar mass in solar masses (column 2), bolometric correction (column 3), 
interstellar absorption (column 4), spectroscopic distance (column 5), distance 
from Hipparcos parallax (column 6), absolute magnitude $M_{\rm V}$ (column 7), 
luminosity (column 8) and the logarithm of the stellar age, 
in years, (column 9) for the barium and the rejected barium stars. Source is the 
same as in Table 1. For stars with ``$^\star$'' see text.}
\begin{tabular}{|l|c|c|c|c|c|c|c|c|c|}\hline
star  & $M/M_{\odot}$ &  B.C. & $A_{V}$ & $r_{\rm spec}$ (pc) & $r_{\rm Hip}$ (pc) & $M_{V}$ & log $L/L_{\odot}$ & $\log t$\,(age) & Source\\\hline
BD-08$^\circ$3194    & 2.0  & $-$0.29  & 0.00 &  366$\pm$86  &   ---       &    1.38 & 1.45 & 9.0 &I\\
BD-09$^\circ$4337    & 1.5  & $-$0.33  & 0.21 &  540$\pm$127 & ---         &    0.21 & 1.94 & 9.5 &\\  
BD-14$^\circ$2678    & 2.5  & $-$0.20  & 0.33 &  488$\pm$115 & ---         &    1.05 & 1.56 & 9.0 &\\  
CD-27$^\circ$2233    & 1.5  & $-$0.38  & 0.18 &  467$\pm$107 & 350$\pm$158 &    0.47 & 1.86 & 9.0 &\\  
CD-29$^\circ$8822    & 3.0  & $-$0.23  & 0.23 &  762$\pm$178 & 427$\pm$272 &    0.21 & 1.90 & 8.7 &\\  
CD-30$^\circ$8774    & 4.0  & $-$0.29  & 0.28 & 1333$\pm$313 & ---         & $-$1.11 & 2.45 & 8.3 &\\  
CD-38$^\circ$585     & 1.0  & $-$0.34  & 0.00 &  540$\pm$127 & ---         &    1.27 & 1.52 &10.0 &\\  
CD-42$^\circ$2048    & 3.0  & $-$0.55  & 0.19 & 1561$\pm$367 & ---         & $-$1.83 & 2.85 & 8.5 &\\  
CD-53$^\circ$8144    & 2.0  & $-$0.33  & 0.24 &  683$\pm$160 & ---         & $-$0.23 & 2.11 & 8.7 &\\  
CD-61$^\circ$1941    & 2.0  & $-$0.33  & 0.51 &  567$\pm$133 & 335$\pm$110 &    0.01 & 2.02 & 9.0 &\\\hline
\end{tabular}  

\par Table 12 is published in its entirety in the electronic edition 
of the Monthly Notices of the Royal Astronomical Society.
A portion is shown here for guidance regarding its form and content.
\end{table*}

\begin{table*}
\caption{Masses of barium stars from this study (second column) and from 
the literature (third column).} 
\begin{tabular}{|l|c|c|c|}\hline                                                          
Star & $M/M_{\odot}$ & $M/M_{\odot}$ & Source \\\hline  
HD 749    & 1.0 & 0.4\,-\,1.0 & Allen \& Barbuy (2006) \\ 
HD 5424   & 1.5 & 1.2\,-\,1.9 & Allen \& Barbuy (2006) \\ 
HD 12392  & 1.5 & 2.0  & Allen \& Barbuy (2006) \\ 
HD 26886  & 3.0 & 2.78($+$0.75,$-$0.78) & Liang et al. (2003) \\  
HD 88562  & 1.5 & 1.0  & Antipova et al. (2004)  \\    
HD 116869 & 2.0 & 0.9\,-\,1.2 & Allen \& Barbuy (2006) \\  
HD 123396 & 1.0 & 0.4\,-\,0.8 & Allen \& Barbuy (2006) \\    
HD 183915 & 4.0 & 3.2  & Antipova et al. (2004)  \\  
HD 201657 & 2.0 & 0.78 & Liu et al. (2009) \\ 
HD 201824 & 3.0 & 4.57 & Liu et al. (2009) \\
HD 204075 & 6.0 & 4.0\,-\,5.0 & B\"ohm-Vitense et al. (2000) \\
          &     & 4.6  & Antipova et al. (2004)  \\
          &     & 4.2  & Smiljanic et al. (2007) \\
HD 210709 & 2.5 & 0.8\,-\,1.1 & Allen \& Barbuy (2006) \\
HD 223938 & 2.0 & 0.6\,-\,1.4 & Allen \& Barbuy (2006) \\\hline
\end{tabular}
\end{table*}

\begin{table*}
\caption{Fit parameters for the distribution of distances to the Galactic plane.
The last column gives the $\chi^{2}$ normalized to the degrees of freedom.} 
\begin{tabular}{|c|c|c|c|c|c|}
\hline 
Fit & $z_{0}$ {[}pc{]} & $\sigma$ {[}pc{]} & $N_{0}$ & $h$ {[}pc{]} & $\chi$$^{2}$$\times$10$^{5}$
\tabularnewline
\hline 
\hline 
Exponential ($z<100$ pc) & --- & --- & $0.40\pm0.06$ & $233\pm24$ & 31.0\tabularnewline
\hline 
Exponential ($z<200$ pc) & --- & --- & $0.21\pm0.03$ & $345\pm61$ & 7.8\tabularnewline
\hline 
Gaussian (all $z$) & $-105\pm28$ & $362\pm28$ & $0.106\pm0.007$ & --- & 22.0\tabularnewline
\hline 
\end{tabular}
\end{table*}

\subsection{Abundances}

\subsubsection{Stars rejected as barium stars}

\par Before discussing the abundance pattern of the barium stars, we
first have to exclude those that display [s/Fe] ratios (Section 4.3)
similar to the field giants that are not s-process enriched.  In
Figure 16, we show the [s/Fe] ratio for the field giants with
metallicities ([Fe/H]) between $-$0.8 and $+$0.5, and the barium star
candidates with similar [s/Fe] ratios of the field giants, including
their error bars. Each error bar represents the mean standard
deviation of the sum of the abundance ratios of [Y/Fe], [Zr/Fe],
[La/Fe], [Ce/Fe], and [Nd/Fe]. Table 15 shows the [s/Fe] ratios for
these stars. We also show that there are five stars (HD 49017, HD
49661, HD 119650, HD 49778 and MFU 214) in Figure 16 (blue and green
squares) with a mean [s/Fe] ratios of $+$0.25, $+$0.26, $+$0.28,
$+$0.29 and $+$0.30, respectively. These stars are located above the
[s/Fe] ratio of the field giants, but considering their error bars
they marginally lie among the field giants. If we assume that the
[s/Fe] ratios of the field giants have the same mean error the barium
stars (0.15 dex), these five stars could not be formally classified as
barium stars, since their [s/Fe] ratios would be practically
indistinguishable from those of the field giants.  Among these five
stars, MFU 214 (=\,BD-01$^\circ$3022, blue square) is a spectroscopic
binary (Jorissen et al. 1998) and therefore is probably a barium star
with a low degree of s-element enrichment. The remaining four stars
with low [s/Fe] ratios, but still above the limit of $~$0.2 dex for
the [s/Fe] ratio of the field giants, could still be labeled as
``probable barium stars''.  In fact, these five stars appear in
Table\,{\sc ii} of MacConnell et al. (1972) as ``Marginal Barium
stars'' and were included in our sample like many others from the same
reference.  Indeed, most of the stars listed in Table 15 belong to
Table\,{\sc ii} of MacConnell et al. (1972).  Two stars not included
in this reference, BD-01$^\circ$302 and HD 211221, but selected from
Gom\'ez et al. (1997), were also rejected as barium stars. These two
stars, however, belong to the list of barium stars of L\"u (1991).

\par There is no clear agreement in the literature on how high should
be the mean [s/Fe] ratio for a star to be considered as a barium star or
even a mild barium star.  Pilachowski (1977) demonstrated that a few
mild barium stars have a mean [s/Fe] of $\geq$+$0.5$, while Sneden et
al. (1981) found $+$0.21. Rojas et al. (2013), analyzed
high-resolution spectra of five ''marginal'' Ba stars from Table\,{\sc
  ii} of MacConnel et al. (1972): CD-65$\degr$2893, HD~22229,
HD~66812, HD~56523, and HD~31341. They derived [s/Fe] ratios of
$+$0.41, $+$0.37, $+$0.35, $+$0.41, and $+$0.34, respectively, which
allowed the authors to classify these stars as ''mild Ba stars''.
Here, we assumed a value of $+$0.25, corresponding to HD 49017, as a
minimum [s/Fe] ratio for a star to be considered a barium star. With
this criterion, we rejected the star HD 95345 (\,=\,58 Leo) as a
barium star because of its low [s/Fe] ratio of $+$0.14. This star has
been already classified as a mild barium star (Sneden et al 1981;
Pinsonneault et al. 1984, and McWilliam \& Lambert 1988) or as a
barium star (Barbuy et al. 1992). Sneden et al. (1981) obtained
$+$0.21 for the [s/Fe] ratio for HD 95345.  The evolutionary status of
mild barium stars is not clear. The authors considered that these
stars were formed in clouds having mild overabundances of the
s-process elements but normal abundances of the other elements.

\begin{figure}
\includegraphics[width=\columnwidth]{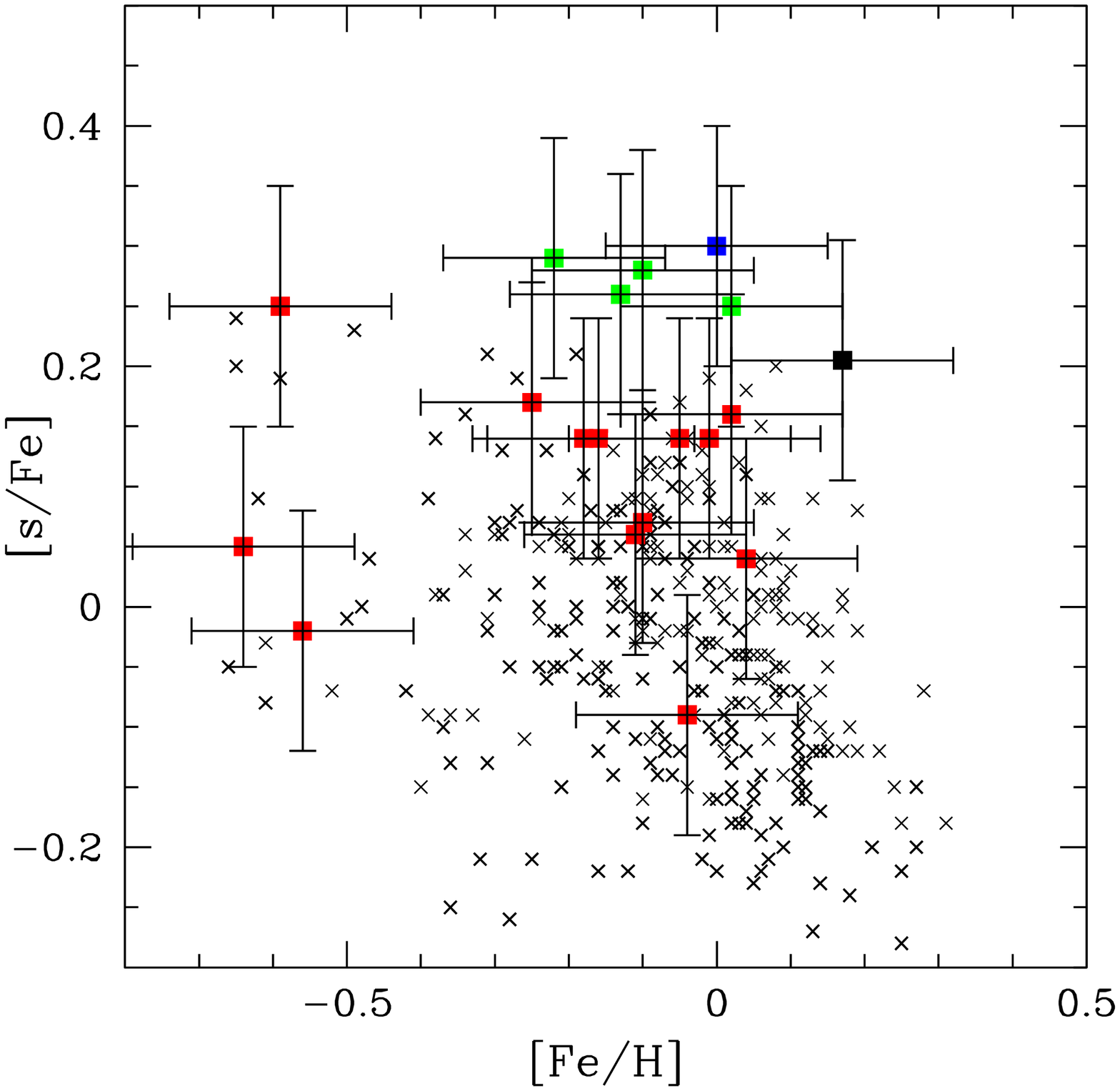}
\caption{[s/Fe] ratios for the rejected barium stars (red squares) in
comparison with field giants (black crosses). Green squares represent
four probable barium stars, HD 49017, HD 49661, HD 119650 and HD 49778. Blue square
represents the star MFU 214 which is a spectroscopic binary.  We also
show in this diagram the position of HD 100012 (black square), the star
rejected previously as a barium star by Pereira et al. (2011).
Data for field giants were taken from Fulbright (2000), Mishenina \& Kovtyukh (2001), 
Mishenina et al. (2006), Luck \& Heiter (2007), Takeda et al. (2008), and 
Sneden et al. (1996).}
\end{figure}

\begin{table}
\centering
\caption{Metallicity and [s/Fe] ratio for the rejected barium stars.}
\begin{tabular}{|c|c|c|}\hline                                        
  star	         &  [Fe/H] & [s/Fe]  \\\hline
BD-01$^\circ$302  &   $-$0.64 &    0.05$\pm$0.21 \\    
HD 5322          &   $-$0.18 &    0.14$\pm$0.12 \\   
HD 21980         &   $-$0.04 & $-$0.09$\pm$0.21 \\    
HD 33409         &   $-$0.59 &    0.25$\pm$0.10 \\   
HD 42700         &   $-$0.25 &    0.17$\pm$0.09 \\    
HD 51315         &   $-$0.05 &    0.14$\pm$0.14 \\    
HD 95345         &   $-$0.16 &    0.14$\pm$0.15 \\    
HD 142491        &   $+$0.02 &    0.16$\pm$0.14 \\
HD 147136        &   $-$0.01 &    0.14$\pm$0.17 \\    
HD 168986        &   $-$0.10 &    0.07$\pm$0.12 \\    
HD 174204        &   $+$0.06 &    0.04$\pm$0.13 \\    
HD 212484        &   $-$0.56 & $-$0.02$\pm$0.11 \\    
HD 211221        &   $-$0.11 &    0.06$\pm$0.17 \\\hline  
\end{tabular}
\end{table}

\subsubsection{Abundance of barium stars\,: Sodium to nickel}

\par In addition to the abundances of the elements created by the
s-process, we also determined the abundances ratios ([X/Fe]) for sodium,
aluminum, $\alpha$-elements (Mg, Si, Ca, and Ti), and iron-peak elements
(Cr and Ni).  Figures 17 and 18 show these abundances ratios for our barium 
giants in comparison with the field giants already considered in
this study (Mishenina et al. 2006; Luck \& Heiter 2007; Takeda et
al. 2008).  We can see that the elements analyzed in this work
(except those of the s-process) follow the same trend observed in the
thin and thick disk giant stars.  The three metal-poor stars analyzed
here, BD+09$^\circ$2384, HD 123396, and HD 130255, also display
an abundance pattern similar to the thick disk/halo stars. For these
three stars, we compared our [X/Fe] ratios with the results of Fulbright (2000),
Mishenina \& Kovtyukh (2001), and Sneden et al. (1996).

\par Sodium and aluminum are mainly produced by hydrostatic carbon
burning in massive stars (Woosley \& Weaver 1995). In the case of
dwarf thin and thick disk stars with metallicities $-$1.0 $<$ [Fe/H]
$<$ $+$0.2, the [Na/Fe] ratio do not display any significant trend
(Edvardsson et al. 1993, Reddy et al. 2003), and even the distinction
between the thin and the thick disk samples based on the [Na/Fe] ratio
is difficult to establish (Reddy et al. 2006). In the case of field
giant stars, both local (Luck \& Heiter 2007) and clump giants
(Mishenina et al. 2006), the [Na/Fe] ratio do not show any trend with
metallicity (Figure 17), like the dwarf stars.  Although many barium
stars have values of the [Na/Fe] ratio similar to the field giants, we
can see several stars in Figure 17 that have an overabundance of the
[Na/Fe] ratio of $+$0.3 to $+$0.6.

\par For the field giants, the [Al/Fe] ratio in this range of
metallicity ($-$0.7 $<$ [Fe/H] $<$ 0.0) was only investigated by Luck
\& Heiter (2007).  Figure 17 shows the behavior of the [Al/Fe] ratio
with metallicity for the barium and field giants. Up to
[Fe/H]\,=\,$-$1.0, the [Al/Fe] ratio increases with decreasing
metallicity (Norris et al. 2001).  Although we do not have stars with
metallicity between $-$1.0 and $-$0.7, both samples (barium giants and
field giants) show a smooth increase of the [Al/Fe] ratio towards
lower metallicity.

\par The [X/Fe] ratio for the $\alpha$-elements for both the barium
stars and the field giants also shows a slight increase with
decreasing the metallicity.  Magnesium is produced in massive stars of
$\sim25\,M_{\odot}$ during the burning of carbon and neon, as
predicted by the nucleosynthesis theory (Woosley \& Weaver 1995). It
would be expected that [Mg/Fe] increases with decreasing metallicity
(Timmes et al. 1995). We can see this behavior in Figure 17, for both
barium and field giants, where the [Mg/Fe] ratio progressively
increases from 0.0 to a maximum value of 0.5--0.6 at [Fe/H] $\sim
-1.5$.

\par Silicon and calcium can be produced by 10-30 $M_{\odot}$ stars by
hydrostatic oxygen burning, and also during the eventual type II
supernovae explosions (Woosley \& Weaver 1986).  For both the field
giants and the barium stars, we observed the same trend for the
[Ca,Si/Fe] ratios as the for the [Mg/Fe] ratio, but with a less sharp
increase.

\par Titanium is usually considered as an $\alpha$-element because its
overabundance in thick disk stars and in metal-poor stars is similar
to that of the other $\alpha$-elements, Mg, Ca, and Si (Fulbright
2000; Reddy et al. 2006; Norris et al. 2001).  Timmes et al. (1995),
however, in their analysis of the chemical evolution of the Galaxy,
included it as an iron-peak element because of its atomic number.
This element is produced by oxygen burning, but can also be produced
by type Ia supernovae (Woosley \& Weaver 1995). Here, we also see the
same trend for both the barium stars and the field giants, i.e. a
slight increase of the [Ti/Fe] ratio with decreasing metallicities.

\par In the case of thin/thick-disk stars, the $\alpha$-elements given
by the mean of Mg, Si, Ca, and Ti are overabundant by $\sim$
0.2\,-\,0.3 dex at $-$1.0 $<$ [Fe/H] $<$ 0.0, and their abundances
then increase up to $\sim$ 0.5 dex at $-$4.0 $<$ [Fe/H] $<$ $-$1.5 for
the halo stars (Edvardsson et al. 1993; Norris et al. 2001).  At
metallicity $-$0.3 to $-$0.5, the thin and thick disks overlap and the
[$\alpha$/Fe] ratio becomes higher for the stars of the thick disk
than for the stars of the thin disk.  At the thick disk, the
[$\alpha$/Fe] ratio may range from to $+$0.2 to $+$0.3 (Reddy et
al. 2006).  Figure 19 shows the [$\alpha$/Fe] ratio {\sl versus}
[Fe/H] for our barium stars in comparison with the local field giant
stars.  We can see that, for an [$\alpha$/Fe] ratio higher than $+$0.2
and in the range of the metallicities between $-$0.6 and $-$0.3, there
are several barium stars with [$\alpha$/Fe] ratio higher than the
local field giants. These barium stars are probably the stars of the
transition of the thin and thick disk, or of the thick disk.  In
Section 5.3, when we analyze the kinematics of barium stars, we
investigate the relationship between the [$\alpha$/Fe] ratio and the
spatial velocities.

\par The iron peak elements such chromium and nickel follow the same
trend of the iron abundance, therefore the [X/Fe] ratios should remain
constant within the range of metallicity studied. In fact, the [X/Fe]
ratios remain constant at [Cr,Ni/Fe]\,=\,0.0 for both field giants and
barium stars.

\par In summary, the abundances ratios of the aluminum,
alpha-elements, and iron peak elements of the barium giant stars
analyzed here are similar to those of the field giants.  When compared
to previous abundance studies of barium stars (Smith 1984; Luck \&
Bond 1991; Liang et al. 2003; Antipova et al. 2003; Jorissen et
al. 2005; Allen \& Barbuy 2006; Smiljanic et al. 2007), our results
for the $\alpha$-elements, as well as for the iron peak elements, are
in good agreement with these studies.  This means that
$\alpha$-elements are slightly enhanced for lower metallicities and
the iron peak elements display no trend in the range of metallicities
we analyzed.  The [X/Fe] ratios of the rejected barium stars do not
present any anomaly in their abundance pattern, as can also be seen in
Figures 17, 18, and 19.

\par Sodium overabundance have been observed in the atmospheres of A-F
supergiant stars (El Eid \& Champagne 1995). According to these
authors, sodium is synthesized in the convective core of main-sequence
stars in the NeNa reaction chain. During the first dredge-up, mixing
brings up the products of the CNO cycle to the star
surface. Therefore, one should expect sodium enrichment in supergiants
and giants rather than in dwarfs.  In fact, Fig.~2 of Boyarchuk et
al. (2001) shows that [Na/Fe] is anti-correlated with $\log g$ and is
higher for $\log g$ {\bf =} 0.0 {\bf --} 0.1 and smaller for $\log g$
{\bf =} 2.0 {\bf --} 3.0.  Sodium production in post-AGB stars was
investigated by Mowlavi (1999).  During the AGB phase, sodium is
synthesized starting from the $^{22}$Ne produced during the He-shell
flash via double-$\alpha$ capture reactions on $^{14}$N (where
$^{14}$N is left as the ashes of H burning), through the reaction
$^{22}$Ne($p$, $\gamma$)$^{23}$Na.  After a thermal pulse, the ashes
of hydrogen burning, including sodium, are mixed and brought up to the
surface (together with $^{12}$C) during the third dredge-up.

\par Antipova et al. (2014) found excess of sodium in some barium
stars in their sample and related this to the evolutionary processes
of these stars, that is the occurrence of the dredge-up of the
produced nuclear-burning material by the convection during the
red-giant stage.  They concluded that, for the giants studied by them,
the [Na/Fe] values do not show significant variations with
metallicity, but the [Na/Fe] ratios were systematically higher for
giant stars with lower values of $\log g$. Here, we also investigated
whether there would be a correlation or anti-correlation between our
[Na/Fe] ratios with $\log g$ for the barium stars. In Figure 20(a), we
can see that there is a trend of increasing [Na/Fe] ratio for
decreasing $\log g$, which is not seen in the samples of Luck \&
Heiter (2007) (Figure 20(c)) and Mishenina et al. (2006) (Figure
20(d)). However, this trend is possibly present in the sample of
Takeda et al. (2008) (Figure 20(b)).  A fit to out data demonstrates
that there is an anti-correlation between the [Na/Fe] ratio and $\log
g$, which can be expressed as
[Na/Fe]$\,=\,(0.39\pm0.05)-(0.09\pm0.02)\,\times(\log g$). This is
shown in Figure 20(a) as a solid line. In addition, we performed two
statistical tests to verify the statistical significance of the
anti-correlation. We determined two correlation coefficients,
$r_{\rm\,P}$ (Pearson test) and $r_{\rm\,S}$ (Spearman test), with
their respective significance level (``prob'') at which the null
hypothesis of zero correlation is disproved (a small value of ``prob''
indicates a significant correlation). We obtained
$r_{\rm\,P}$\,=\,$-$0.33 and $r_{\rm\,S}$\,=\,$-$0.35, with $prob
(r)_{\rm\, P}\,=\,1.9\times10^{-5}$ and $prob
(r)_{\rm\,S}\,=\,0.5\times10^{-5}$. These values indicate that there
exist a weak but statistically significant anti-correlation between
[Na/Fe] ratio and $\log g$.

\par Since sodium can be produced at the post-AGB phase, as already
explained, we may consider that sodium-rich material was transferred
to the star during the mass-loss at the end of the AGB phase, and the
star also becomes enriched in the elements created by the
s-process. But prior to the AGB, the surface of the star may be
enriched in sodium.  A star with at least 1.5\,$M_{\odot}$ is able to
raise the sodium abundance through the NeNa cycle (Denissenkov \&
Ivanov 1987) in its hydrogen-burning core while it is still on the
main sequence. Later, when the star becomes a giant, the first
dredge-up brings the synthesized sodium to the stellar
surface. Therefore, we should also consider the possibility that a
contribution to the sodium overabundance seen in Figure 20 might come
from the barium star itself.

\begin{figure} 
\includegraphics[width=\columnwidth]{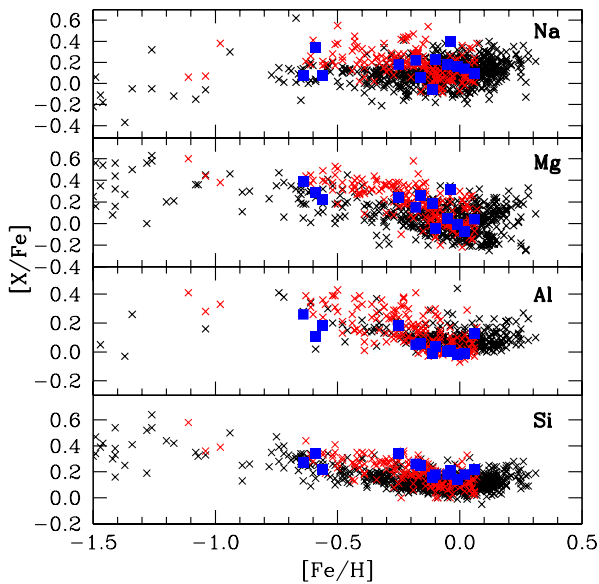}
\caption{Abundance ratios [X/Fe] {\sl versus} [Fe/H] for Na, Mg, Al and Si. 
Barium giants analyzed in this work ({\it red crosses}), 
rejected barium stars ({\it blue squares}), and field giants ({\it black crosses}). 
Data for field giants are the same as in Figure 16.}
\end{figure}

\begin{figure}
\includegraphics[width=\columnwidth]{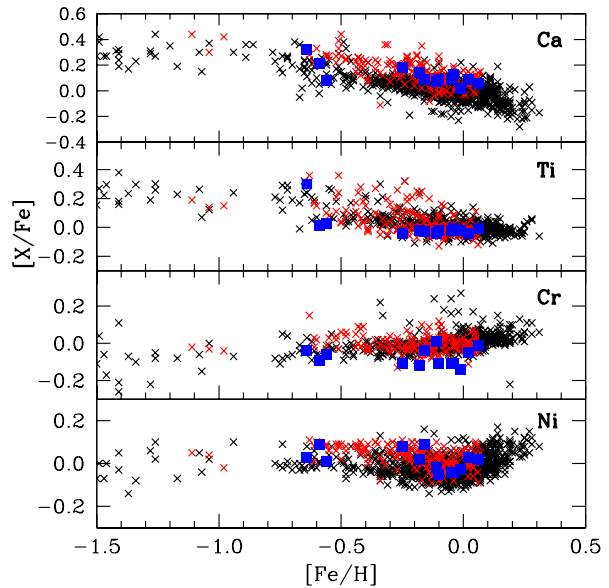}
\caption{Abundance ratios [X/Fe] {\sl versus} [Fe/H] for Ca, Ti, Cr and Ni.
 Symbols have the same meaning as in Figure 17. Data for field giants are the 
same as in Figure 16.}
\end{figure}

\begin{figure}
\includegraphics[width=\columnwidth]{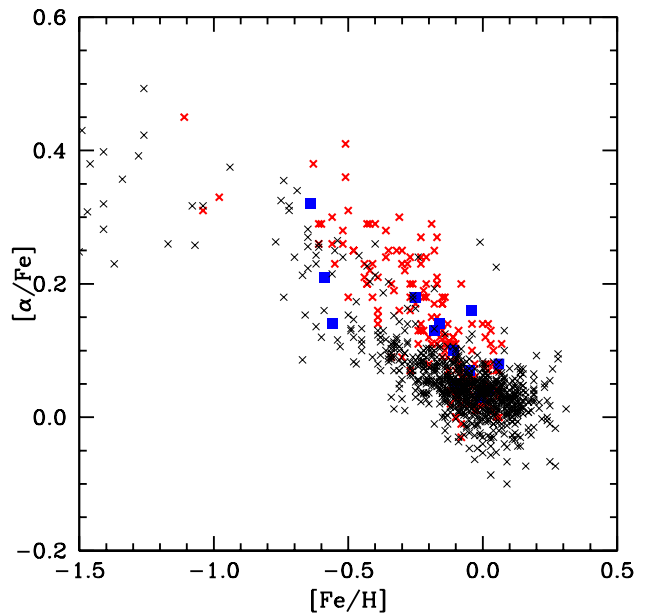}
\caption{Abundance ratio [$\alpha$/Fe] {\sl versus} [Fe/H]. Symbols have the 
same  meaning as in Figure 17. Data for field giants are the same as in Figure 16.}
\end{figure}

\begin{figure} 
\includegraphics[width=\columnwidth]{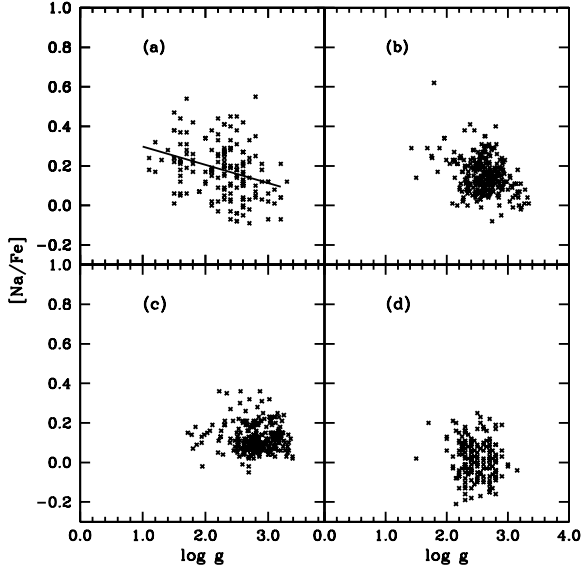}
\caption{[Na/Fe] ratio {\sl versus} $\log g$ for the barium giants (a). 
We also show the behavior of the [Na/Fe] ratio with $\log g$ 
for the samples of Takeda et al. (2008) (b), Luck \& Heiter (2007) (c) and 
Mishenina et al. (2006) (d). In (a) we show the fit for the relation between 
the [Na/Fe] ratio and the $\log g$ for the barium giants.}
\end{figure}

\subsubsection{Abundance of barium stars\,: Heavy-element abundance}

\par Figure 21 shows the [X/Fe] ratios for the elements created by the
s-process\,: Y, Zr, La, Ce, and Nd, for the barium giants analyzed in
this work (red crosses) in comparison with the field giants (green
crosses).  As already discussed in Section 5.2.1, we considered barium
stars as those with enhancement factors [s/Fe] of at least
$+$0.26. This is the case of HD 49017, although some stars may have an
[X/Fe] ratio lower than this value for some given element.  We can see
in Figure 21 that the abundance of zirconium is poorly investigated in
normal giant stars in this metallicity range.  Barium stars with
metallicity higher than $+$0.10 analyzed by Pereira et al. (2011), and
the stars HD 10613, with a metallicity of [Fe/H]\,=\,$-$0.81, and HD
206983, with a metallicity of [Fe/H]\,=\,$-$1.4, previously analyzed
by Pereira \& Drake (2011) and Pereira \& Junqueira (2001), were
included in Figure 21.  The figure also shows other determinations of
the [Y, Zr, La, Ce and Nd/Fe] ratios for barium giants (filled black
squares) and barium dwarfs (open black squares) taken from the
literature.

\par Models of galactic chemical evolution do not predict the observed
overabundances of the s-process elements seen in Figure 21 in this
range of metallicity (Travaglio et al. 1999, 2004). Therefore the
atmospheres of these stars were contaminated either by any intrinsic
process such as self-enrichment or by an extrinsic event that may have
happened in the past, i.e., the mass-transfer hypothesis.  The first
hypothesis can be ruled out because as seen in Figure 12, their
luminosities are too low to be an AGB star and therefore become
self-enriched.

\par Figure 22 shows the [s/Fe] and the [hs/ls] ratios {\sl versus}
metallicity for the same data points of Figure 21, together with other
classes of chemically peculiar binary systems, such as the yellow
symbiotic stars (blue squares, Pereira \& Roig 2009), the CH stars
(blue stars, van Eck et al. 2003), and five CEMP-s (carbon enhanced
metal-poor) stars that have already been proved to be binary systems
(red circles): CS~22948-027, CS~29497-030, CS~29497-034, CS~22964-161,
and HE~0024-2523. The data of these binary CEMP-s concerning their
carbon and heavy-element (Z $>$ 56) overabundances were taken from
Preston \& Sneden (2001), Sivarani et al. (2004), Barbuy et
al. (2005), Lucatello et al. (2003), Thompson et al. (2008).  We also
added in Figure 22 the [s/Fe] ratio for some post-AGB stars (green
filled circles) with data taken from Reyniers et al. 2004 and Pereira
et al. 2012. The abundance of the s-elements of the post-AGB stars is
mainly similar to that of the barium stars, although in some cases may
be higher. This is expected since barium stars owe their overabundance
of the s-process elements to the previous AGB phase.

\par From Figures 21 and 22, we can see that there is no distinction
between giant and dwarf barium stars, as far as the abundance of the
elements of the s-process is concerned.  In the sample of the
main-sequence turnoff CEMP-s stars analyzed by Aoki et al. (2008),
these authors showed that the [C/H] ratios are higher than the same
ratio in giant stars. This led the authors to conclude that no
significant dilution have happened in these stars, which means that
thermohaline mixing was not very efficient.  On the other hand, Husti
et al. (2009) showed that the effects of dilution factors should be
considered in their analysis and interpretation of the abundances of
the dwarf barium stars. They considered that, once barium dwarfs do
not have a deep convective envelope, dilution factors are mandatory to
fit the theoretical predictions to the observed [X/Fe] ratios. This
has been taken as an evidence that thermohaline mixing is an efficient
process, that must be taken into account in the analysis of the dwarf
barium stars.

\par Among the elements created by the s-process nucleosynthesis, the
first neutron magic number peak elements, i.e. the light elements
([ls]) such as Y and Zr, are the dominant products of the neutron
captures in AGB models at metallicities higher than solar (Busso et
al. 2001).  According to these models, negative values for the [hs/ls]
ratio means that the elements Y and Zr are more abundantly produced
for moderate neutron exposures at these metallicities.  At lower
metallicities, [Fe/H]$\sim$$-$1.0, the second neutron magic number
peak elements of the s-process, i.e. the heavy elements ([hs]) such as
Ba-La-Ce-Nd, are the dominant. Therefore, the [hs/ls] is a useful
measure of the neutron capture efficiency and has been widely used in
the AGB nucleosynthesis models.

\par The [hs/ls] ratio has its maximum value of $\sim$$+$0.6 around
metallicities between $\sim$$-$0.6 and $\sim$$-$0.4 (Figure 14 of
Goriely \& Mowlavi 2000).  According to stellar models given by Busso
et al. (2001), the maximum of [hs/ls]\,=\,1.2 takes place at a
metallicity $-$1.0, depending on the amount of $^{13}$C burnt.  At
even lower metallicities the first-peak elements are bypassed in favor
of the second-peak elements and those of the third peak. Therefore, we
should expect that the lead abundance tends to be higher for lower
metallicity objects than for most of the barium stars analyzed in this
work, which is actually observed for the CH stars (van Eck et
al. 2003; Pereira \& Drake 2009). This is due to the neutron exposure
that increases with decreasing number of seed nuclei or
metallicity. AGB models predict the values of the [hs/ls] ratio for
stars of different metallicities.  In Figure 14 of Goriely \& Mowlavi
(2000), the authors show how this ratio behaves with the number of
thermal pulses. Busso et al. (2001) also provides the run of the
[hs/ls] ratio with metallicity for a 1.5\,$M_{\odot}$ and a
3.0\,$M_{\odot}$ AGB star and for different choices of $^{13}$C
pocket. In Figure 12 of Cristallo et al. (2011), the nucleosynthesis
models include the influence of the stellar mass on the [hs/ls]
ratio. From the above references, we can see the different ways that
the stellar masses and/or the number of thermal pulses are related to
the [hs/ls] ratio, but the same general behavior with the metallicity
is found in all the models.  The negative values of the [hs/ls] ratio
are expected to be seen for metallicities higher than $-$0.2 (Figure
14 of Goriely \& Mowlavi 2000).

\par The behavior of the [hs/ls] ratio and also the [s/Fe] ratio {\sl
  versus} metallicity, which increase with decreasing metallicity, is
a consequence of the operation of the reaction
$^{13}$C($\alpha,\,n$)$^{16}$O, since this neutron source is
anti-correlated with metallicity (Clayton 1988; Wallerstein 1997).
When a large sample of barium stars is investigated, we can detect
this anti-correlation.  We found that, while most of the barium stars
have metallicities in the range $-$0.6 $\le$[Fe/H]$\le$0.2, the CEMP-s
stars have [Fe/H] $\le$ $-$2.0 and the CH stars have intermediate
metallicities ($-$1.5 $\le$[Fe/H]$\le$$-$0.5), which are in the range
where the barium stars are rare. Moreover, there is not a clear
separation among these three groups of chemically peculiar stars in
terms of metallicity.

\par In Figure 23, we show the [hs/ls] ratio {\sl versus} the [s/Fe]
ratio for the barium stars analyzed in this work. There is a clear
correlation between these two ratios. A linear least-squares fit gives
a correlation coefficient of $+$0.79 and the fit $[{\rm
    hs/ls}]\,=\,(-0.30\pm0.03)+(0.62\pm0.04)\times[{\rm s/Fe}]$
(excluding the rejected barium stars, indicated by blue squares in
this Figure). The same kind of trend was observed in post-AGB stars
that display the spectral feature at 21\,$\mu$m (Figure 7 of Reyniers
et al. 2004). The 21\,$\mu$m feature carrier has been observed in the
spectra of carbon-rich post-AGB stars, that is, objects with a C/O
ratio close or greater than 1.0. Although its origin has not been
identified yet, several suggestions for this feature have been
discussed in the literature.  These bands are generally accepted to be
due to polycyclic aromatic hydrocarbons (see Tielens 2008 for a
review).  The observed trend of the [hs/ls] and [s/Fe] ratios would
reflect the fact that a high neutron efficiency implies a very
efficient dredge-up process. However, as highlighted by Busso et
al. (2001), the behavior of these ratios should display a looplike
trend, as can be seen in their Figures 11 and 12, and this would be a
consequence of dilution factors of the s-process matter. In fact, in
post-AGB stars, Reyniers et al. (2004) obtained a higher correlation
coefficient ($+$0.96) than ours ($+$0.79).  Since barium stars owe
their s-process abundances to the former AGB star, which is not the
case for the intrinsic post-AGB stars, s-process rich material was
mixed and diluted into the envelope of the future barium star.  The
differences between the post-AGB and barium stars given by the
different fits between the [hs/ls] and [s/Fe] ratios is a consequence
of the different dilution factors during the transfer of mass to the
atmosphere of the future barium star.  In addition, the spread (or the
scatter) seen in the [s/Fe] ratio is also related to other parameters
of the barium stars, such as the orbital period and eccentricity of
the binary system, the metallicity, the number of thermal pulses, and
the initial mass.

\par Figure 23 also separates the barium stars of our sample according
to their metallicities. The stars marked in magenta represent those of
metallicities between $+$0.3 and 0.0, stars marked with blue crosses
have metallicities between 0.0 and $-$0.4, stars marked with red
squares have metallicities between $-$0.4 and $-$0.5, and stars marked
with green squares have metallicities between $-$0.5 and $-$1.4. The
reason for this division is because the efficiency of the neutron
capture given by the [hs/ls] ratio is useful in the comparison of this
observed ratio with the same ratios predicted by AGB models. In fact,
it has been shown that this ratio is metallicity dependent, as can be
seen in Figures 3a and 4a of Busso et al. (2001) and in Figure 14 of
Goriely \& Mowlavi (2000).

\par Most of the stars marked with green squares in this diagram have
positive [hs/ls] ratios, except HD 139409 and HD 134698 that have a
[hs/ls] ratio of $-$0.05 and $-$0.12, respectively.  HD 139409 was
analyzed before by Za$\check{\rm c}$s (1994) and Antipova et
al. (2004).  Antipova et al. (2004) also obtained a low [hs/ls] ratio
of $+$0.09, with a metallicity similar to us: $-$0.51. Za$\check{\rm
  c}$s (1994) obtained a higher [hs/ls] ratio of $+$0.24, that could
be due to a different surface gravity.  HD 134698 also has a low
[hs/ls] ratio and a metallicity of $-$0.52.  The observed [hs/ls]
ratios for these two stars may be considered atypical when compared to
the models of Goriely \& Mowlawi (2000), since these models predicted
negative [hs/ls] values only for metallicities down to $\sim$$-$0.1.
The stars with this metallicity may also have negative [hs/ls] values
according to the models of Busso et al. (2001).

\par Among the red square in Figure 23, HD 4084 has the lowest [hs/ls]
ratio, 0.04. HD 4084 was previously analyzed by Barbuy et al. (1992)
but the authors did not obtain the heavy s-element abundances.

\par Blue crosses, which represent the majority of the stars of our
sample of studied barium stars (116 stars), occupy all parts of the
diagram along the fit.  Among these, some stars have
high values of the [hs/ls] ratio, considering their ``high''
metallicities.  Specifically, CPD-64$^\circ$4333 with
[Fe/H]\,=\,$-$0.10 and [hs/ls]\,=\,0.65; HD 12392 with
[Fe/H]\,=\,$-$0.08 and [hs/ls]\,=\,0.65 (Allen \& Barbuy 2006 obtained
for this star [Fe/H]\,=\,$-$0.06 and [hs/ls]\,=\,0.63); HD 24035 with
[Fe/H]\,=\,$-$0.23 and [hs/ls]\,=\,0.68; HD 92626 with
[Fe/H]\,=\,$-$0.15 and [hs/ls]\,=\,0.88 and HD 123949 with
[Fe/H]\,=\,$-$0.07 and [hs/ls]\,=\,0.71.

\par Stars in magenta are located in the region below [hs/ls] $<$ 0.25
since all of them have metallicities between 0.0 and $-$0.35.  Among
the magenta points, we also found stars with probably high values of
the [hs/ls] ratio considering their metallicities. The stars
CD-29$^\circ$8822, HD 49017, HD 182300 and HD 204886, with
metallicities of $+$0.02, $+$0.02, $+$0.06, and $+$0.04, respectively,
have a mean [hs/ls]\,=\,0.26$\pm$0.03. This behavior is similar to
that of the star HD 46040, analyzed in Pereira et al. (2011), which
has [Fe/H]\,=\,0.11 and a high [hs/ls] ratio of 0.29.  Finally, we
show in Figure 23 the positions of HD 204075 and HD 221879. These
stars are massive stars and good candidates to be objects in which the
neutrons were released by the reaction $^{22}$Ne($\alpha$,n)$^{25}$Mg
(see below).

\par Since we can consider the masses of the barium stars as a lower
limit for the AGB stars responsible for their chemical peculiarities,
is it possible that the s-elements distribution in barium stars would
also be a result of nucleosynthesis in AGB stars of different
masses\,?  In order to try to answer this question, we investigated
whether there is a possible relationship between the mass of barium
star and the abundance of s-process elements using the [s/Fe] and the
[hs/ls] ratios. Figure 24 shows the dependence of [s/Fe] and [hs/ls]
on stellar mass. For stars with masses of 1.0, 1.5, 2.0, 2.5, 3.0,
4.0, 5.0 and 6.0 $M_{\odot}$ we found a mean [s/Fe] ratios of,
respectively, 1.07$\pm$0.45 (8 stars), 1.12$\pm$0.39 (12 stars),
0.96$\pm$0.29 (42 stars), 0.84$\pm$0.24 (12 stars), 0.78$\pm$0.34 (64
stars), 0.84$\pm$0.37 (27 stars) 0.43$\pm$0.19 (5 stars) and
0.74$\pm$0.22 (3 stars), respectively. As far as the [s/Fe] ratio is
concerned, we can see in Figure 24 a wide range of [s/Fe] ratios
values for a given mass. Among the stars with 1.0 and 4.0 $M_{\odot}$
there is no statistically significant difference among the [s/Fe]
ratios, taking into account the corresponding [s/Fe] uncertainties.

\par For stars with 5.0 $M_{\odot}$ there is a decrease in the [s/Fe]
ratio which could possibly indicate a low efficiency in the production
of the s-process elements in the previous AGB star.  Although based on
the results of only three stars, the higher value of the [s/Fe] ratio
for stars with 6.0 $M_{\odot}$ (0.74), which is close to the value for
the stars of 3.0 $M_{\odot}$ and 4.0 $M_{\odot}$, strengths the
possible origin in the previous AGB star and would not be related to
the nature of the neutron source.  The higher values for [s/Fe] ratios
for the stars of lower masses (stars of 1.0, 1.5 and 2.0 $M_{\odot}$)
should be taken with caution, because not only the [s/Fe] ratio is
strongly anti-correlated with metallicity but also because some
low-metallicity giants are found among low mass stars (Jorissen et
al. 1998).

\par Contrary to the [s/Fe] ratio, the [hs/ls] ratio may probably aid
this discussion if, for example, we were able to distinguish which was
the neutron source in the former AGB star. We found a mean of [hs/ls]
ratio of 0.47$\pm$0.22, 0.35$\pm$0.31, 0.35$\pm$0.25, 0.16$\pm$0.19,
0.18$\pm$0.25, 0.25$\pm$0.26, 0.01$\pm$0.06 and $-$0.18$\pm$0.09 for
the stars from 1.0 to 6.0 $M_{\odot}$ respectively.  For the [hs/ls]
ratio does not show any statistically significant difference between
stars of 1.0, and 4.0\,$M_{\odot}$.  The higher values of the [hs/ls]
ratio seen for stars of 1.0, 1.5 and 2.0 $M_{\odot}$ can not be
attributed only to the difference in mass solely. Like the [s/Fe]
ratio (Figure 22), the [hs/ls] is anti-correlated with metallicity.
The [hs/ls] ratio as also displays a wide range of values for a given
mass as well.  This means that other parameters such as the number of
thermal pulses of the previous AGB star, the efficiency of thermal
pulses, dilution factors, metallicity, orbital separation, the way the
matter was transferred from the AGB star (wind accretion in a detached
binary system or by Roche lobe overflow) among others (Jorissen et
al. 1998; Liang et al. 2000; Busso et al. 2001, Lugaro et al. 2004)
are important to define the chemical peculiarities observed in the
barium stars.

\par However, it is interesting to note that the stars with 5.0
$M_{\odot}$ and 6.0\,$M_{\odot}$ have the lowest [hs/ls] ratios among
all the mass range.  These stars are HD 58121, HD 119650, HD 148177,
HD 176105, and HD 210030 (5.0 $M_{\odot}$) and HD 204075, HD 216809
and HD 221879 (6.0 $M_{\odot}$). Three of five stars with 5.0
$M_{\odot}$ also have low [s/Fe] ratios (HD 58121 with
[hs/ls]\,=\,0.01 and [s/Fe]\,=\,0.34; HD 119650 with
[hs/ls]\,=\,$-$0.01 and [s/Fe]\,=\,0.28; and, HD 210030 with
[hs/ls]\,=\,$-$0.08 and [s/Fe]\,=\,0.31). This means that these stars
have received a low s-process enriched material, both the light
s-process elements (Y and Zr) and the heavy s-process elements (La,
Ce, and Nd). The other two stars with 5.0 $M_{\odot}$ show mild
s-process enrichment.

\par Among the other three stars with 6.0 $M_{\odot}$, two of them
display high [s/Fe] ratios and low [hs/ls] ratios: HD 204075 with
[s/Fe]\,=\,0.96 and [hs/ls]\,=\,$-$0.24; and HD 221879, with
[s/Fe]\,=\,0.75 and [hs/ls]\,=\,$-$0.22.  The remaining one, HD
216809, only shows moderate enrichment with [s/Fe]\,=\,0.52 and a low
ratio [hs/ls]\,=\,$-$0.08.  Of these stars, only HD 204075 had its
abundance pattern investigated before. Antipova et al. (2003) and
Smiljanic et al. (2007), respectively, determined [hs/ls]\,=\, $-$0.04
and $-$0.06 while we obtained $-$0.24 using the same elements. The
abundances of the heaviest s-process elements (from Gd to Pb) in this
star were obtained by Gopka et al. (2006).

\par Low [hs/ls] ratios are expected for stars with masses less than
$M \leq (3\,-4)\,M_{\odot}$ at near solar metallicities, considering
the reaction $^{13}$C($\alpha,\,n$)$^{16}$O as the neutron
source. However, more massive AGB stars, between 5.0 $M_{\odot}$ and
8.0 $M_{\odot}$, can also display low [hs/ls] ratios, considering the
reaction $^{22}$Ne($\alpha$,n)$^{25}$Mg as the most likely neutron
source for the s-process (Busso et al. 2001; Karakas \& Lattanzio
2014).  Karakas \& Lattanzio (2014) (see also Karakas et al. 2012)
predicted that, for stars of 5.0 and 6.0 $M_{\odot}$, the [X/Fe]
ratios of the elements heavier than Z$>$30, the elements of the first
peak of the s-process (Rb, Sr, Y, and Zr), are predominantly produced
over the elements of the second peak of the s-process (Ba-La-Ce-Nd).
this has been taken as an evidence of the operation of the reaction
$^{22}$Ne($\alpha$,n)$^{25}$Mg.  In fact, an AGB model for a 5.0
$M_{\odot}$ star and solar metallicity given by Karakas et al. (2012)
predicts a mild overall s-process enhancements, compatible with the
[s/Fe] and [hs/ls] ratios observed in the 6.0 $M_{\odot}$ star HD
216809 and probably also for the other three 5.0 $M_{\odot}$ stars HD
58121, HD 119650 and HD 210030. All these massive stars are the
natural candidates to have companions that were even more massive so
that the reaction of $^{22}$Ne($\alpha$,n)$^{25}$Mg happened as the
source of free neutrons.  In HD 204075 and HD 221879 the mean
abundance of the second peak of the s-process (La, Ce and Nd)
determined in this work (and also in other studies for HD 204075) are
higher when compared with the AGB models.

\par Karakas \& Lattanzio (2014) raised the importance of mass,
besides metallicity in the yields of AGB stars. For barium stars with
5.0 $M_{\odot}$ and 6.0 $M_{\odot}$, the [hs/ls] ratios at solar
metallicity make them good candidates to investigate the stellar
yields in stars of intermediate mass. We found that several barium
stars of our sample have 4.0 $M_{\odot}$. Therefore, we may ask how
much mass was transferred from their companions, assuming that if the
companions were more massive than 4.0 $M_{\odot}$ in the past.  Boffin
\& Jorissen (1988), for example, estimated that one solar mass can be
accreted by the secondary.  Barium stars with 4.0 $M_{\odot}$ display
a mean [hs/ls] ratio of 0.30$\pm$0.22 which is higher than the mean
seen for the stars of 5.0 $M_{\odot}$ and 6.0 $M_{\odot}$. According
to the AGB models in Karakas et al. (2012) their companions were not
an intermediate mass stars, otherwise we would probably have much
lower [hs/ls] ratios. If this is true, therefore the mass accreted by
these 4.0\,$M_{\odot}$ stars was not much higher than
1.0\,$M_{\odot}$.  Therefore, our determined [hs/ls] ratios they may
be useful to identify which was the neutron source in the former AGB
star, and this in turn would be related to the mass (and metallicity)
of these stars.  Up to 4.0 $M_{\odot}$, the most likely neutron source
is still $^{13}$C, while in stars with 5.0 and 6.0 $M_{\odot}$, it is
$^{22}$Ne.

\begin{figure} 
\includegraphics[width=\columnwidth]{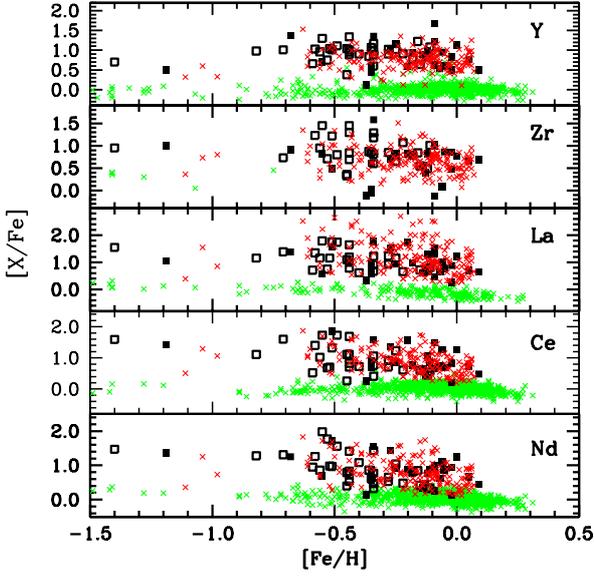}
\caption{Abundance ratio [X/Fe] {\sl versus} [Fe/H] for the elements of the
s-process. Green crosses represent field giants and red crosses the barium
giants analyzed in this work. Filled and open squares represent the [X/Fe] ratios, 
respectively, for barium giants and dwarfs from the literature.}
\end{figure}

\begin{figure} 
\includegraphics[width=\columnwidth]{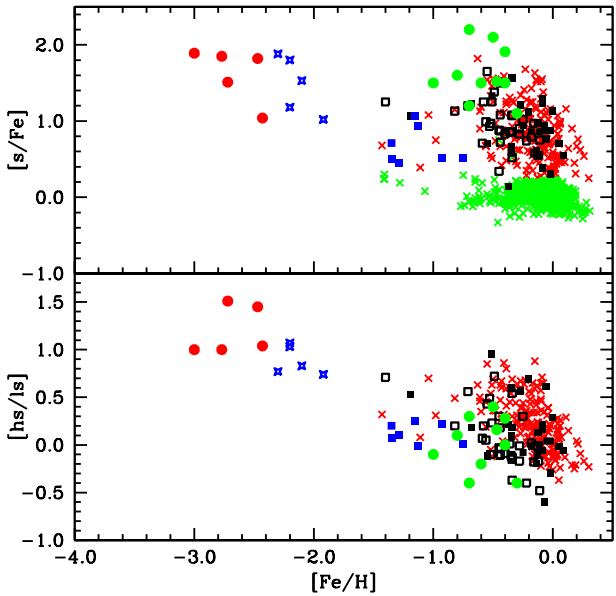}
\caption{Diagram of [s/Fe] {\sl versus} [Fe/H] (top) and [hs/ls] {\it versus} 
[Fe/H] (bottom) for several classes of chemically peculiar binary stars.
For field giants and barium giants analyzed in this work, symbols have the same 
meaning as in Figure 21.
CEMP-s stars which are member of binary systems are represented by
{\it red filled circles}, CH stars by {\it blue stars}, yellow symbiotics
by {\it blue filled squares} and post-AGB stars by {\it green filled circles}
Filled and open black squares are the same as in Figure 23.}
\end{figure}

\begin{figure} 
\includegraphics[width=\columnwidth]{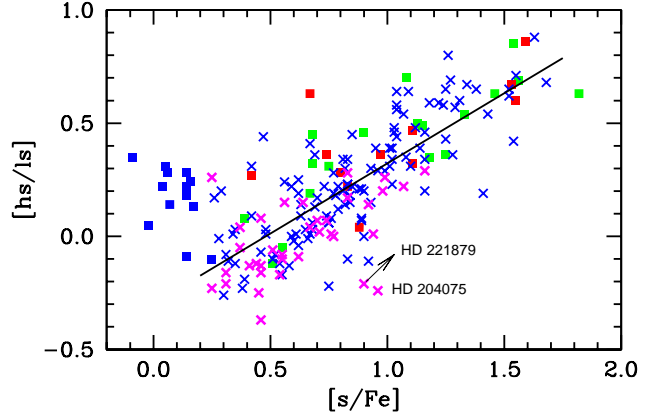}
\caption{Correlation between [hs/ls] and [s/Fe] ratios for the barium stars 
analyzed in this work. The straight line provides the least-square fit for the
whole sample. In this Figure we distinguish barium stars according to their 
metallicities. The stars with magenta colors represent those with metallicities 
between $+$0.3 and 0.0; stars with blue crosses, between 0.0 and $-$0.4; stars 
with red squares, between $-$0.4 and $-$0.5, and stars with green colors between 
$-$0.5 and $-$1.4. Blue squares represent the rejected barium stars and they 
occupy the same region where the field giant stars would be in this diagram.} 
\end{figure}

\begin{figure} 
\includegraphics[width=\columnwidth]{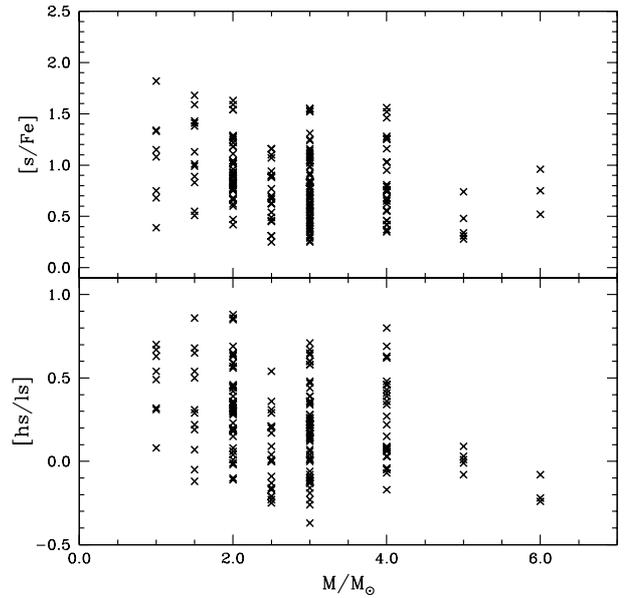}
\caption{[s/Fe] and [hs/ls] ratios {\sl versus} stellar mass for the barium giants.} 
\end{figure}

\subsection{Kinematics}

\par To analyze the kinematical properties of our sample of barium
giants, we used the distances that were calculated in Section 5.1 and
the radial velocities that were determined by measuring the Doppler
shift of the spectral lines.  We did not determine the space
velocities for HD 43389, HD 66291, HD 74950 and HD 252117 because we
could not determine the interstellar extinction in the direction to
these stars, nor their distances. Yet, we obtained their radial
velocities, which are 50.3$\pm$0.2 km\,s$^{-1}$, 21.8$\pm$0.1
km\,s$^{-1}$, 13.0$\pm$0.2 km\,s$^{-1}$, and 25.8$\pm$1.3,
respectively. Due to the same reason, we did not determine the
distances of the rejected barium stars HD 142491 and HD 168986 either,
but their radial velocities are $-$12.8$\pm$0.8 km\,s$^{-1}$ and
13.0$\pm$0.2 km\,s$^{-1}$, respectively.

\par We obtained the spatial velocities relative to the local standard
of rest, $U_{\rm LSR}$, $V_{\rm LSR}$, $W_{\rm LSR}$, where $U_{\rm
  LSR}$ is positive toward the Galactic center, $V_{\rm LSR}$ is
positive in direction of Galactic rotation ($l$\,=\,90$^\circ$,
$b$\,=\,0$^\circ$), and $W_{\rm LSR}$ is positive toward the north
Galactic pole ($b$\,=\,90$^\circ$). We assumed the solar motion of
(11.1, 12.2, 7.3) km\,s$^{-1}$, as derived by Sch\"onrich et
al. (2010).  For these calculations, we employed the algorithm of
Johnson \& Soderblom (1987) using proper motions from Perryman et
al. (1997) and H\o g et al. (1998).  As we did in previous Sections,
we also compared our results to the space velocities of the field
giants. Takeda et al. (2008) already provided the spatial velocities
for their sample, while for the sample of Luck \& Heiter (2007) we
calculated the velocities based on the distances and the radial
velocities given in that paper.  For the sample of Luck \& Heiter
(2007), we obtained mean values of $U_{\rm LSR}= -$9.34 km\,s$^{-1}$,
$V_{\rm LSR}= -$17.6 km\,s$^{-1}$, and $W_{\rm LSR}= -$9.0
km\,s$^{-1}$, that can be compared to $-$9.4 km\,s$^{-1}$ , $-$17.5
km\,s$^{-1}$, and $-$9.2 km\,s$^{-1}$ reported by those authors.  We
also calculated the probability for a barium star to be a member of
the thin disk, the thick disk, or the halo population, following the
procedure described in Reddy et al. (2006).  Membership to a given
population is established when the star has a probability $P_{\rm
  'population'}$ greater than or equal to 70\%.

\par Table 16 shows the results obtained for the spatial velocities
and the corresponding probabilities. As seen in this Table, the
majority of the barium stars belong to the thin disk population, but
some other stars can be members of the thin/thick disk, the thick
disk, or even the halo population.  Of the 178 stars surveyed in the
kinematical analysis, 160 stars (90\% of the sample) belong to the
thin disk population.

\par Figure 25 shows the Toomre diagram of $(U^{2}_{\rm LSR} +
W^{2}_{\rm LSR})^{1/2}$ {\sl versus} $V_{\rm LSR}$, where the stars
are kinematically classified according to their spatial velocities and
probabilities. Red crosses represent the barium stars of the thin
disk, green crosses represent the barium stars in the transition
thin/thick disk, and blue crosses are the barium stars of the thick
disk. Magenta crosses represent the barium stars in the halo.  Among
the four halo stars, three of them (BD+09$^\circ$2384, HD 10613 and HD
206983) fulfill the basic requisites to be considered halo objects:
low metallicity, high spatial velocities, and enhanced [$\alpha$/Fe]
ratios, while the remaining one, HD 221879, although kinematically
considered as a halo object, it maybe a thick disk star (see
discussion ahead in this Section).

\par Another relevant information about a given stellar population in
the Galaxy is provided by the mean spatial velocities and the
dispersions in these velocities ($\sigma_{\rm U}$, $\sigma_{\rm V}$,
$\sigma_{\rm W}$).  Table 17 shows the results for the mean spatial
velocities and dispersions of the barium giants analyzed in this work
and the field giant stars from the samples of Luck \& Heiter (2007)
and Takeda et al. (2008).  In addition, we provide the dispersions in
the spatial velocities for the different populations in the Galaxy,
that were used to constrain the probability for a barium stars to
belong to a given population.  Table 17 shows that the two samples of
field giants analyzed by Luck \& Heiter (2007) and Takeda et
al. (2008) belong to the same population, that is, the thin disk.
G\'omez et al. (1997) and Mennessier et al. (1997) also analyzed a
large sample of barium stars from L\"u (1991).  Their results for
group 2 of G\'omez et al. (1997) and for group G (giants and
subgiants) of Mennessier et al. (1997) are the most numerous and are
shown in Table 17. These authors found dispersion velocities similar
to ours for the thin disk population.

\par The presence of a halo component among the barium stars was
identified in the kinematical analysis of Gom\'ez et al. (1997) and
Mennessier et al. (1997).  Three stars kinematically classified by
Gom\'ez et al. (1997) as members of halo population (HD 10613, HD
104340 and HD 206983) agreed with spectroscopic studies showing that
these stars are also metal-poor objects (Junqueira \& Pereira 2001;
Pereira \& Drake 2009). According to Table 16, the stars HD 88927, HD
115277 and HD 219116 are objects of the thin disk population. For HD
107541, our results indicate a probability of only 74\% of being a
member of the thin disk, and it could be considered a star in the
transition of the thin to the thick disk.  Mennessier et al. (1997)
had some stars in common with our study.  These authors also
recognized that HD 10613, HD 104340, and HD 206983 are members of the
halo population.  HD 187762 behaves kinematically like HD 107541,
having a probability of 77\% of being a member of the thin disk.
CD-27$^\circ$2233 and HD 5424 are also members of the thin disk
population.  Our kinematical analysis shows that HD 139409 is a member
of the thick disk rather than the halo population.

\par Figure 26 shows the dependence of the metallicity and the
     [$\alpha$/Fe] ratio on the spatial velocity ($V_{\rm SPA}$),
     where $V_{\rm SPA}\,=\,(U^{2}_{\rm LSR} + V^{2}_{\rm LSR}$ $+$
     W$^{2}_{\rm LSR}$)$^{1/2}$, for the barium giants analyzed in
     this work. This Figure provides support to our previous
     conclusions about the population distribution of barium
     stars. Stars with lower spatial velocities and higher
     metallicities have a trend to be members of the thin disk, while
     stars with higher spatial velocities and lower metallicities have
     a trend to be members of the thick disk and halo.  In Figure 26,
     there are seven stars that deserve separate discussion.
     BD+09$^\circ$2384, HD 10613, HD 206983 (magenta crosses) are halo
     stars already mentioned.  HD 221879 (magenta cross) has a high
     spatial velocity of 207.9 km\,s$^{-1}$, which is in the border of
     the values of the spatial velocities of the thick disk and halo
     according to data from Reddy et al. (2006).  Our analysis showed
     that HD 221879 has a probability of 77\% to be a member of the
     halo, however it has a metallicity and [$\alpha$/Fe] ratio
     typical of the thin disk stars. It is a massive star, with 6.0
     $M_{\odot}$, and hence it is unlikely to be a member of the halo
     or the thick disk. Therefore, HD 221879 can be considered as a
     ``kinematically thick\,-\,metallicity thin disk'' star. Dwarf
     stars with these properties have been already found by Reddy et
     al. (2006) in their abundance and kinematical analysis of thick
     disk stars. Mishenina et al. (2004) also found some dwarf stars
     displaying this paradoxal behavior. These stars were probably
     originated in the thin disk and later were collisionaly scattered
     into the thick disk (Reddy et al. 2006).

\par In Figure 26 there are seven stars that deserve a
comment. BD+09$^\circ$2384, HD 10613, HD 206983 (magenta crosses) are
halo stars, already above mentioned.  HD 221879 (magenta cross) has a
high spatial velocity of 207.9 km\,s$^{-1}$ which is in the border of
the values of the spatial velocities of the thick disk and halo using
data from Reddy et al. (2006).  Our probability calculation shows that
HD 221879 has a probability of 77\% to be a member of the halo,
however it has a metallicity and the [$\alpha$/Fe] ratio typical to
the thin disk stars. It is a massive star, with 6.0 $M_{\odot}$, and
hence unlikely to be a member of the halo or the thick disk. Therefore
HD 221879 can be considered as a ``kinematically thick\,-\,metallicity
thin disk'' star. Dwarf stars of these kind have already been found in
a study done by Reddy et al. (2006) in their abundance and kinematical
analysis of thick disk stars. Mishenina et al. (2004) also found some
dwarf stars displaying this paradox behavior. Probably these stars
have originated in the thin disk and later were collisionaly scattered
into the thick disk (Reddy et al. 2006).

\par HD 123396 (green cross in Figure 26) is kinematically classified
as a transition object between the thin to the thick disk, but it is a
low metallicity object ([Fe/H]\,=\,$-$1.04). It is probably a thick
disk object, with a probability of 64\%. HD 130255 (blue cross), with
a metallicity of [Fe/H]$\sim$$-$1.1, could be a candidate for the halo
population, but it has been kinematically classified by us as a thick
disk star with 92\% probability. This is not surprising since stars of
the thick disk may have metallicities up to $\sim$$-$1.1 (Robin et
al. 2003), although stars with much lower metallicities ($\sim$$-$2.2)
have been identified as probable members of the thick disk (Beers et
al. 2002).

\par HD 119185 (open square in Figure 26) has a probability of 66\% to
be a member of the halo and, in view of this, it was classified as an
object in the transtion between the thick disk and the halo.  Its high
distance from the Galactic plane ($-$1.0 kpc) combined with its high
spatial velocity (243.7 km\,s$^{-1}$) indicate that this star could be
a halo star.  However, its metallicity ([Fe/H]\,=\,$-$0.43) and its
[$\alpha$/Fe] ratio ($+$0.18) do not allow us to classify HD 119185 as
a halo star.

\par Finally, Table 18 shows the kinematic data for the rejected
barium stars. We may notice that most of them belong to the thin
disk. One star, HD 212484 (open black square in Figure 26), has
probability to belong to the thick disk/halo population. Another one,
BD-01$^\circ$302 (magenta square in Figure 26), has probability to
belong to the halo. This star was also classified by G\'omez et
al. (1997) as a halo star, in agreement with our results.

\begin{figure} 
\includegraphics[width=\columnwidth]{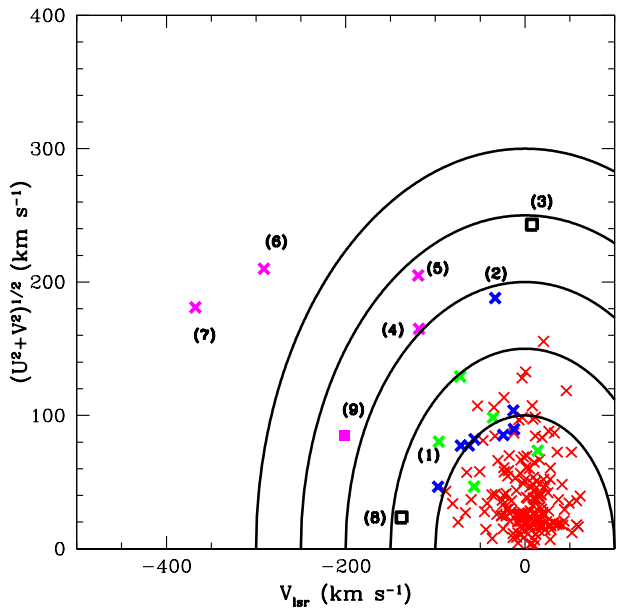}
\caption{Toomre diagram for the barium stars analyzed in this work.
Red crosses represent the stars of the thin disk population; green
the stars of the thin/thick and blue the stars of the thick disk. 
Magenta crosses represent the stars of the halo population. The 
numbers represent the stars HD 123396 (1); HD 130255 (2); HD 119185 (3);
HD 221879 (4); HD 10613 (5); BD-09$^\circ$2384 (6); HD 206983 (7).
The numbers (8) and (9) represent, respectively, two rejected barium stars
HD 212484 and BD-01$^\circ$302.}
\end{figure}

\begin{figure} 
\includegraphics[width=\columnwidth]{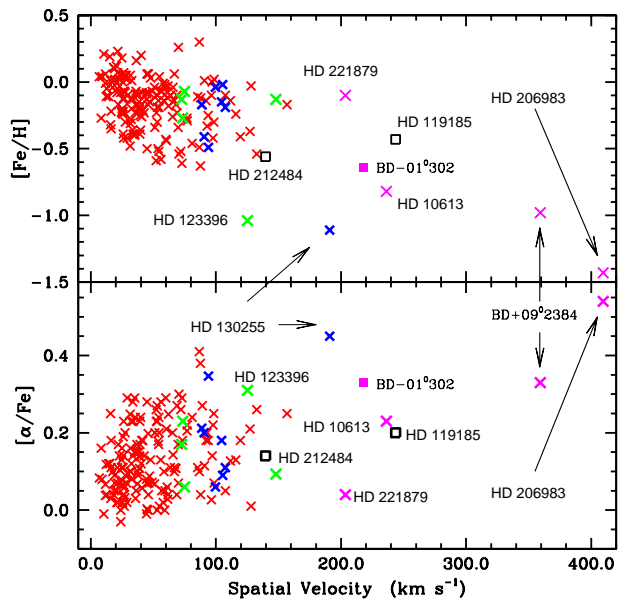}
\caption{Metallicity and [$\alpha$/Fe] ratio dependence {\sl versus} spatial
velocity for the barium and the rejected barium stars analyzed in this work. 
Symbols have the same meaning as in Figure 25. 
In this Figure we also show some the metal-poor barium stars BD+09$^\circ$2384,
HD 10613,  HD 123396, HD 130255 and HD 206983 and two stars which were
kinematically classified as a member of thick disk/halo and halo,
HD 119185 and HD 221879, respectively. Two rejected barium stars, HD 212484
(open square) kinematically classified as a member of thick disk/halo 
and BD-01$^\circ$302 (magenta square) kinematically classified as a member of the 
halo are also shown.}
\end{figure}

\begin{table*} 
\caption{Kinematic data for the true barium stars. The radial velocities are given
in the second column, the spacial velocities with their respective uncertainties are given from 
the third to the eigth columns. The ninth, tenth and eleventh columns give the probability 
of a star being a member of the thin disk ($P1$), thick disk ($P2$) and the halo ($P3$). 
The last column gives the spatial velocity of the star.}
\begin{tabular}{cccccccccccc}\hline
Star &  $RV$ &  $U_{\rm LSR}$ & $eU$ & $V_{\rm LSR}$ & $eV$ & $W_{\rm LSR}$ & $eW$ & $P1$ & $P2$ & $P3$ & $V_{\rm SPA}$\\\hline
     &  km\,s$ ^{-1}$ & km\,s$ ^{-1}$ &  & km\,s$ ^{-1}$ &  & km\,s$ ^{-1}$ &  & & & & km\,s$ ^{-1}$ \\\hline
\multicolumn {6}{c}{\bf {thin disk stars}} \\\hline
BD-08$^\circ$3194  & 4.2$\pm$0.5     & $-$25.0  & 3.3  &  10.9   & 0.9 & 11.2   & 0.8  & 1.00 &  0.00 &  0.00 & 29.5 \\
BD-09$^\circ$4337  & $-$36.5$\pm$0.4 &    25.4  & 1.2  & $-$3.4  & 3.1 & $-$0.1 & 2.3  & 1.00 &  0.00 &  0.00 & 25.6 \\
BD-14$^\circ$2678  & 3.6$\pm$ 0.4    & $-$18.2  & 2.5  &   2.1   & 2.0 &  2.5   & 1.9  & 1.00 &  0.00 &  0.00 & 18.5 \\
BD-18$^\circ$821   & 36.0$\pm$0.2    &   36.5   & 6.1  &$-$21.0  & 4.7 & 19.6   & 8.6  & 0.98 &  0.02 &  0.00 & 46.5 \\
CD-26$^\circ$7844  & 13.7$\pm$0.1    & $-$24.1  & 3.7  & $-$8.8  & 2.2 & $-$4.1 & 4.3  & 1.00 &  0.00 &  0.00 & 26.0 \\
CD-27$^\circ$2233  & $-$8.1$\pm$0.3  & $-$75.0  & 12.0 & $-$46.9 & 13.0&  31.2  & 4.1  & 0.79 &  0.21 &  0.00 & 107.4 \\
CD-29$^\circ$8822  & 14.7$\pm$0.4    & $-$3.2   & 2.9  & $-$7.7  & 1.9 &  3.0   & 3.1  & 1.00 &  0.00 &  0.00 &  8.9 \\
CD-30$^\circ$8774  & $-$1.5$\pm$0.1  &    5.9   & 5.1  &  10.5   & 1.6 & 3.0    & 3.0  & 1.00 &  0.00 &  0.00 & 12.5 \\   
CD-30$^\circ$9005  &$-$11.8$\pm$0.1  &    23.0  & 7.9  &  13.6   & 2.3 & $-$5.3 & 2.4  & 1.00 &  0.00 &  0.00 & 27.3 \\     
CD-34$^\circ$6139  & 1.3$\pm$0.1     &   45.1   & 13.2 &   9.6   & 0.9 &$-$16.4 & 6.2  & 1.00 &  0.00 &  0.00 & 48.9 \\\hline
\end{tabular}

\par Table 16 is published in its entirety in the electronic edition 
of the Monthly Notices of the Royal Astronomical Society.
A portion is shown here for guidance regarding its form and content.
\end{table*}

\begin{table*} 
\caption{Our results for the mean spatial velocities $\langle$$U_{\rm LSR}$$\rangle$, 
$\langle$$V_{\rm LSR}$$\rangle$  and $\langle$$W_{\rm LSR}$$\rangle$ with their respective 
dispersions in these velocities
($\sigma$$_{U}$, $\sigma$$_{V}$, $\sigma$$_{W}$) for the barium stars (BaS) 
analyzed in this work. The spatial velocities and their respective dispersions 
for two samples of field giants as well as one of the results from the statistical 
analysis of barium stars of G\'omez et al. (1997) and Mennessier at al. (1997) are 
also shown. At the end of the Table we provide the dispersions in the spatial 
velocities of the different populations in the Galaxy such as the thin disk, 
thick disk and halo population.}
\begin{tabular}{ccccc}\hline                  
Sample & $\langle$$U_{\rm LSR}$$\rangle$ (km\,s$^{-1}$) & $\langle$$V_{\rm LSR}$$\rangle$ (km\,s$^{-1}$) 
& $\langle$$W_{\rm LSR}$$\rangle$ (km\,s$^{-1}$) & \# stars \\
       & $\sigma$$_{U}$ & $\sigma$$_{V}$ & $\sigma$$_{W}$ & Ref.\\\hline

Thin disk BaS     &  6.2 &  1.6  &  0.3 &  160 \\
                  & 45.2  & 27.6  & 19.8 &  This work  \\\hline

Thin disk/thick disk BaS &  63.5  & $-$49.3 &   8.4 & 5  \\
                         &    57.3  &    41.8 &  41.1 & This work \\\hline

Thick disk BaS     & $-$7.3 & $-$46.3 & $-$28.7 &  8 \\
                   &  81.2  &    30.5 &    64.6 &  This work \\\hline

Barium giants      & $-$10.0 & $-$13.0  & $-$7.0  & 159 \\
                   &    36.0 &    20.0  &    16.0 & G\'omez et al. (1997) \\\hline

Barium giants      & $-$14.0 & $-$12.0  & $-$6.0  & 159 \\
                   &    37.0 &    20.0  &    16.0 &  Mennessier et al. (1997) \\\hline

Field giants       &  $-$9.3  & $-$17.6 & $-$9.0 & 298 \\
                   &    34.7  &    25.0 &   16.9 & Luck \& Heiter (2007) \\\hline

Field giants       &     1.1  &    3.2  &  $-$0.8 & 322 \\
                   &    32.1  &   23.1  &    15.8 & Takeda et al. (2008)  \\\hline\hline
       & $\sigma$$_{U}$ & $\sigma$$_{V}$  & $\sigma$$_{W}$ & \\\hline
Thin disk          &  43    &  28  &  17  &  Reddy et al. (2006) \\\hline
Thick disk         &  67    &  51  &  42  &  Reddy et al. (2006) \\\hline
Halo               & 131    & 106  &  85  &  Reddy et al. (2006) \\\hline
\end{tabular}
\end{table*}

\begin{table*} 
\caption{Kinematic data for the rejected barium stars. The radial velocities are given
in the second column, the spacial velocities are given in the third, fourth and
fifth column. The sixth, seventh and eighth columns give the probability of star being a member
of the thin disk ($P1$), thick disk ($P2$) or the halo ($P3$). The last column gives
the total spatial velocity of the star.}
\begin{tabular}{cccccccccccc}\hline
Star &  $RV$ &  $U_{\rm LSR}$ & $eU$ & $V_{\rm LSR}$ & $eV$ & $W_{\rm LSR}$ & $eW$ & $P1$ & $P2$ & $P3$ & $V_{\rm SPA}$\\\hline
     &  km\,s$ ^{-1}$ & km\,s$ ^{-1}$ &  & km\,s$ ^{-1}$ &  & km\,s$ ^{-1}$ &  & & & & km\,s$ ^{-1}$ \\\hline
\multicolumn {5}{c}{\textbf {thin disk stars}} \\\hline
HD 5322    &  $-$1.9$\pm$0.2  &  $-$1.5  & 3.0 &    17.9  & 2.3  & 3.7     & 1.7 & 1.00 & 0.00 & 0.00 & 18.3 \\ 
HD 21980   &    69.1$\pm$0.1  &    14.3  & 3.5 & $-$68.6  & 10.7 & $-$17.8 & 7.5 & 0.89 & 0.11 & 0.00 & 72.3 \\
HD 33409   &    37.1$\pm$0.1  &  $-$41.2 & 7.9 &  $-$0.7  & 4.2  & $-$40.4 & 6.3 & 0.96 & 0.04 & 0.00 & 57.6 \\
HD 42700   &     5.4$\pm$0.1  &     16.4 & 6.6 &     10.1 & 1.3  & 7.7     & 2.1 & 1.00 & 0.00 & 0.00 & 20.7 \\
HD 51315   &    12.3$\pm$0.1  &     18.7 & 8.3 &     10.7 & 3.1  & $-$2.4  & 2.6 & 1.00 & 0.00 & 0.00 & 21.6 \\
HD 95345   &     7.4$\pm$0.2  &  $-$24.3 & 3.4 &      2.1 & 1.5  & 12.9    & 0.4 & 1.00 & 0.00 & 0.00 & 27.6 \\
HD 147136  & $-$25.4$\pm$0.2  & $-$9.6   & 1.0 &     22.8 & 2.9  & $-$8.2  & 2.1 & 1.00 & 0.00 & 0.00 & 26.0 \\
HD 174204  &     1.9$\pm$0.1  &     15.2 & 6.7 &  $-$54.3 & 15.6 & $-$5.7  & 4.6 & 0.96 & 0.04 & 0.00 & 56.6 \\ 
HD 211221  & $-$18.4$\pm$0.1  &     40.1 & 9.5 &  $-$14.3 & 5.8  & $-$8.6  &  7.1& 0.99 & 0.01 & 0.00 & 43.2 \\\hline
\multicolumn {6}{c}{\textbf {thick disk/halo }} \\\hline
HD 212484     &  45.3$\pm$0.1   & 20.1  & 13.1 & $-$137.6 & 29.7 & 12.4 & 8.5 & 0.11 & 0.37 & 0.51 & 139.6 \\\hline
\multicolumn {6}{c}{\textbf {halo star}} \\\hline
BD-01$^\circ$302 & $-$32.1$\pm$0.2 & $-$77.8 & 13.7 & $-$200.9 & 50.9 & $-$35.0 & 17.3 & 0.00 & 0.00 &  1.00 & 218.2\\\hline
\end{tabular}
\end{table*}

\section{Conclusions}

\par Based on high-resolution optical spectroscopy data, we determined
the atmospheric parameters and abundances of Na, Al,
$\alpha$-elements, iron-peak elements, and the elements created by the
s-process for a large sample of barium stars.  Our spectroscopic
analysis indicates that the distributions of metallicity, temperature,
and surface gravity cannot be represented by one single gaussian
distribution.  The metallicity distribution showed that barium stars
have one major peak at [Fe/H]\,=\,$-$0.12 and another at
[Fe/H]\,=\,$-$0.49. These two metalicity mean values are similar to
those of the thin disk and the thick disk, respectively, according to
results reported in the literature. The observed distribution of
surface gravity for barium stars can be fit by three gaussians, thus
indicating the presence of evolved giants among the barium stars
population.  The temperature distribution can be fit by two gaussians,
one corresponding to the majority of the barium giants and another
representing the evolved and cooler population of barium stars.  The
determination of the surface gravity and temperature allowed us to
obtain the masses and ages of the barium stars, and to determine the
relation between age and metallicity, and between age and mass. Barium
giant stars also follow the age-metallicity relation, i.e. younger
objects are more metal rich than the older ones. They also follow the
same trend between mass and age observed for the field giants by
Takeda et al. (2008), i.e. younger stars are more massive while older
ones are less massive.

\par Based on the ionization equilibrium we determined the surface
gravities and, hence, the spectroscopic distances which enable us to
determine the luminosities and scale height of the barium stars. We
concluded that none of the barium giants is luminous enough to be an
AGB star and become self-enriched in the s-process elements. Our
determination of the scale height shows that barium stars have a scale
height similar to the disk stars.

\par The abundances of aluminum, $\alpha$-elements and iron-peak
elements of barium stars are similar to the field giants of similar
metallicities.  We found that some barium stars have higher sodium
abundance compared to the field giant stars. This conclusion is
strengthened by the anti-correlation between the [Na/Fe] ratio and
surface gravity, that is, stars with lower surface gravity tend to be
more sodium rich than stars with higher surface gravities. This is
probably due to the production of sodium in the former AGB star (now
the white dwarf) of the binary system that transferred sodium enriched
matter and/or that the barium star itself might become self enriched
in sodium due to the NeNa cycle.

\par After determining the abundance of the elements created by the
s-process for the 182 stars of our sample, we disregarded the ``barium
star nature'' of 13 stars since they present a mean [s/Fe] ratio
similar to the non-s-process enriched field giant stars. Our abundance
analysis showed that heavy-element abundance pattern of barium stars
showed different degrees of enrichment, considering the [s/Fe] ratio
as a diagnostic of s-process enrichment. We did not find a distinction
in the [s/Fe] ratio between our sample of barium stars and the barium
giants previously analyzed in the literature. We did not find a
difference of the [s/Fe] of barium stars analyzed in this paper and
that of dwarf barium stars.

\par The [s/Fe] and the [hs/ls] ratios increase as the metallicity
decreases, as expected based on the theoretical models of
nucleosynthesis in AGB stars considering $^{13}$C as the neutron
source.  Our [hs/ls] ratio has a maximum value of $\sim$$+$0.5 at a
metallicity of $\sim$$-$0.5.  Using the diagram [hs/ls] {\sl versus}
[s/Fe] ratio and dividing the stars according to their metallicities,
we found that some stars may have atypical [hs/ls] ratios considering
their metallicities.  The massive stars of our sample, stars with 5.0
$M_{\odot}$ and 6.0 $M_{\odot}$, are probably stars where the source
of neutrons is $^{22}$Ne by the reaction
$^{22}$Ne($\alpha$,n)$^{25}$Mg.

\par Based on the determined spectroscopic distances, the measurements
of radial velocities, and using the proper motions from the
literature, we determined the $U_{\rm LSR}$, $V_{\rm LSR}$ and $W_{\rm
  LSR}$ with their respective dispersions, as well as the spatial
velocities ($V_{\rm SPA}$) of our sample of barium giants and the
samples of field giants analyzed by Luck \& Heiter (2007) and Takeda
et al. (2008).  This analysis allowed us to obtain the probability of
a star to be a member of a given population in the Galaxy.  Our
results showed that 90\% of the barium stars belong to the thin disk,
while for the field giants from the samples of Luck \& Heiter (2007)
and Takeda et al. (2008) the fraction is 98\%. For the thin/thick disk
and the thick disk populations, the barium stars proportionally have a
much bigger population, 3.0\% and 4.0\%, compared to 1.0\% and 0.7\%
for the field giants. For the thick disk/halo and halo populations,
the fraction of barium stars is 3.0\% compared to only 0.3\% of the
field giants.  In addition, combining the results on spatial
velocities, metallicities and [$\alpha$/Fe] ratios, we showed that
barium stars follow the expected behavior in which stars with lower
spatial velocities are in the thin disk while stars with higher
spatial velocities are in the thick disk and the halo. Metal-poor
barium stars analyzed in this work also have kinematical properties of
the thick disk/halo and halo populations.

\par Finally, we would like to stress that our work is far from being
completed.  The abundances of light elements (carbon, nitrogen,
oxygen, and lithium), as well as the $^{12}$C/\,$^{13}$C isotopic
ratio, were not analyzed here and will be a matter of further
investigation.  Carbon and nitrogen are very strongly affected by
nuclear burning in the inner regions of the stars, such as the CNO
cycle and the triple-$\alpha$ reaction.  When the convection in the
stellar envelope penetrates inwards during the stellar evolution along
the giant branch, the material which has been processed in the
interior is mixed to the surface. Therefore, the determination of the
abundances of carbon, nitrogen, lithium, and the $^{12}$C/\,$^{13}$C
isotopic ratio is essential for understanding not only convection, but
also to set observational constraints to the nucleosynthesis history
of the stars and the current theory of stellar evolution.

\par Rubidium, lead, and europium are another interesting abundance
targets.  The abundance of rubidium can provide another distinction,
besides the [hs/ls] ratio, between the high mass and the low mass
barium stars.  In addition, detailed nucleosynthesis models for stars
with masses between 5.0 $M_{\odot}$ and 9.0 $M_{\odot}$ at solar
metallicity predict [Rb/(Sr,Zr)]$>$0.0 (Karakas et al. 2012). This is
also a subject for future study, since it implies to analyze a large
sample of the low mass barium giant stars, as well as to compare them
with the high mass barium stars, in order to have a statistically
significant sample. Lead abundance is also important to investigate
the contribution of the strong component of the s-process in a sample
of chemically peculiar stars covering a wide range of metallicities.

\par As seen in Figure 22, the abundance of the s-process elements of
barium stars is similar to that of some post-AGB stars.  Europium in
some post-AGB stars presents a very high [Eu/Fe] ratio ($+$1.55 in
IRAS 06530$-$0213, and $+$1.03 in IRAS 08143$-$4406, Reyniers et
al. 2004; $+$ 1.33 in GLMP 334, and $+$1.01 in IRAS 15482$-$5741,
Pereira et al. 2012), but some post-AGB stars also show low [Eu/Fe]
ratio as, for example, $+$0.26 in IRAS 19500$-$1709 (Van Winckel \&
Reyniers 2000).  Europium is 94\% formed by the r-process (Arlandini
et al. 1999), but in post-AGB stars is formed by the s-process
(Pereira et al. 2012).  Therefore, the abundance of europium in a
sample of barium stars will be important to check whether a similar
high [Eu/Fe] ratios, like those found in post-AGB stars, are detected
in barium stars.

\section{Acknowledgements} 

\par This research has made use of the SIMBAD database, operated at
CDS, Strasbourg, France. N.A.D. acknowledges FAPERJ, Rio de Janeiro,
Brazil, for Visiting Researcher grant E-26/200.128/2015 and the Saint
Petersburg State University for research grant 6.38.18.2014. E. Jilinski
acknowledges the PCI program under the grant 454794/2015-0.

{}

\bsp
\label{lastpage}
\end{document}